\documentclass[lettersize,journal]{IEEEtran}
\usepackage{amsmath,amsfonts}
\usepackage{algorithmic}
\usepackage{algorithm}
\usepackage{array}
\usepackage[caption=false,font=normalsize,labelfont=sf,textfont=sf]{subfig}
\usepackage{textcomp}
\usepackage{stfloats}
\usepackage{url}
\usepackage{verbatim}
\usepackage{graphicx}
\usepackage{booktabs}
\usepackage{cite}
\usepackage{bm}

\usepackage{tikz}
\usepackage{pgfplots}
\usepgfplotslibrary{groupplots}
\usetikzlibrary{positioning}
\usepackage{acronym}

\newcommand{\LatVec}{\mathbf{y}}
\newcommand{\LatVecQ}{\hat{\mathbf{y}}}
\newcommand{\SQVar}{z}
\newcommand{\SQVarQ}{\hat{z}}
\newcommand{\SQVec}{\mathbf{z}}
\newcommand{\SQVecQ}{\hat{\mathbf{z}}}

\newcommand{\STFTVar}{S}
\newcommand{\STFTVarRec}{\hat{S}}
\newcommand{\STFT}{\mathbf{S}}
\newcommand{\STFTRec}{\hat{\STFT}}
\newcommand{\XVec}{\mathbf{x}}
\newcommand{\XVecRec}{\hat{\XVec}}
\newcommand{\NoiseVec}{\bm{\epsilon}}
\newcommand{\NoiseVar}{\epsilon}

\newcommand{\Var}{\sigma^2}

\newcommand{\BatchSize}{B}
\newcommand{\NumCh}{C}
\newcommand{\FrameIdx}{n}
\newcommand{\NumFrames}{N}
\newcommand{\FreqIdx}{f}
\newcommand{\NumFreqs}{F}
\newcommand{\Dilation}{D}
\newcommand{\Stride}{S}
\newcommand{\KernelSize}{K}
\newcommand{\NumBlocks}{M}
\newcommand{\ConvLin}{\mathtt{Conv1\hspace{-3pt}\times\hspace{-3pt}1}}
\newcommand{\CdbkIdx}{q}
\newcommand{\CdbkIdxVec}{\mathbf{q}}
\newcommand{\NumLevels}{Q}
\newcommand{\QuantIdx}{r}
\newcommand{\NumQuants}{R}
\newcommand{\QuantLevel}{\Delta}
\newcommand{\DimData}{D}
\newcommand{\DimLatent}{d}

\newcommand{\SetIndVQ}{\mathcal{Q}_\text{VQ}}
\newcommand{\SetIndSQ}{\mathcal{Q}_\text{SQ}}

\newcommand{\Expect}{\mathbb{E}}
\newcommand{\KLD}{\mathcal{D}_{\text{KL}}}
\newcommand{\EOP}[2]{\Expect_{#1}\left[{#2}\right]}
\newcommand{\KLDOP}[2]{\KLD\left({#1}\Vert {#2}\right)} 

\newcommand{\transp}{^{\text{T}}}
\newcommand{\Encoder}{\mathtt{enc}}
\newcommand{\Decoder}{\mathtt{dec}}
\newcommand{\Qenc}{\mathtt{Q}_{\text{enc}}}
\newcommand{\Qdec}{\mathtt{Q}_{\text{dec}}}

\acrodef{SQ}{Scalar Quantization}
\acrodef{VQ}{Vector Quantization}
\acrodef{BN}{BatchNorm}
\acrodef{DNN}{Deep Neural Network}
\acrodef{GAN}{Generative Adversarial Network}
\acrodef{MSE}{Mean Square Error}
\acrodef{NN}{Neural Network}
\acrodef{VAE}{Variational Autoencoder}
\acrodef{RVQ}{Residual VQ}
\acrodef{PDF}{Probability Density Function}
\acrodef{STFT}{Short-Time Fourier Transform}
\acrodef{LS}{Least Squares}

\newcommand{\Revision}[1]{#1}

\begin{document}

\title{\Revision{Neural Speech Coding\\ for Real-time Communications\\ using Constant Bitrate Scalar Quantization}}

\author{Andreas Brendel,~\IEEEmembership{Member,~IEEE}, Nicola Pia, Kishan Gupta, Lyonel Behringer, Guillaume Fuchs, Markus Multrus
\thanks{The authors are with the Fraunhofer IIS, Fraunhofer Institute for Integrated Circuits IIS in Erlangen, Germany. Contact: andreas.brendel@iis.fraunhofer.de}
}

\markboth{Journal of \LaTeX\ Class}%
{Brendel \MakeLowercase{\textit{et al.}}: A Sample Article Using IEEEtran.cls for IEEE Journals}


\maketitle

\begin{abstract}
	Neural audio coding has emerged as a vivid research direction by promising good audio quality at very low bitrates unachievable by classical coding techniques. Here, end-to-end trainable autoencoder-like models represent the state of the art, where a discrete representation in the bottleneck of the autoencoder \Revision{is learned. This} allows for efficient transmission of the input audio signal. \Revision{The learned discrete representation of neural codecs is} typically generated by applying a quantizer to the output of the neural encoder. In almost all state-of-the-art neural audio coding approaches, this quantizer is realized as a Vector Quantizer (VQ) and a lot of effort has been spent to alleviate drawbacks of this quantization technique when used together with a neural audio coder. In this paper, we propose \Revision{and analyze} simple alternatives to VQ, which are based on projected Scalar Quantization (SQ). These quantization techniques do not need any additional losses, scheduling parameters or codebook storage thereby simplifying the training of neural audio codecs. \Revision{For real-time speech communication applications, these neural codecs are required to operate at low complexity, low latency and at low bitrates. We address those challenges by proposing a new causal network architecture that is based on SQ and a Short-Time Fourier Transform (STFT) representation. The proposed method performs particularly well in the very low complexity and low bitrate regime.}
\end{abstract}

\begin{IEEEkeywords}
Discrete representation learning, \Revision{low complexity}, neural \Revision{speech} coding, quantization, \Revision{real-time}
\end{IEEEkeywords}

\section{Introduction}
\IEEEPARstart{S}{peech} is the most natural form of human communication and, hence, \Revision{technology for the transmission of speech waveforms have ever been important and has continuously improved over the past decades.} Here, the needs of a constantly increasing number of users has to be served by enabling efficient communication with limited shared resources at the same time. Therefore, compressing the transmitted waveforms in order to minimize transmission datarate\Revision{, algorithmic latency and computational complexity} while maintaining speech quality is still a key research problem. 

In classical waveform-matching speech coding, many advanced signal processing techniques are used to remove redundancy and perceptually irrelevant components from the \Revision{signals \cite{oshaughnessy_review_2023}}. Such approaches produce \Revision{very good} natural speech quality, depending on the bit rate, but cannot be extended to very low bit rates, \Revision{below $4-6\,\mathrm{kbps}$,} where the paradigm fails. \Revision{Conventional} parametric coding techniques can be used at very low bit rates and still produce intelligible speech, but of poor audio quality. While classic, signal processing-based audio coding \Revision{is} still an active field of research, in recent years speech and audio coding methods based on \acp{DNN} have gained a lot of attention from the research community. \Revision{One of the  main reasons for this is the} good quality at very low bitrates \Revision{achieved by neural codecs}, which is challenging for classical methods.
\subsection{Neural Speech and Audio Coding}
Codecs relying on \acp{DNN} can be roughly classified into three categories: Neural post filters, neural vocoders and end-to-end approaches.
\subsubsection{Neural Post Filters}
\Revision{Neural post filters can enhance the} decoded speech signal of a classical codec on receiver side, which enables backward compatibility with existing and legacy technologies. Hence, neural post filtering approaches have the advantage of being \Revision{easy to integrate} into existing systems. Following this paradigm, a convolutional \ac{DNN} \cite{zhao_convolutional_2019} and a masked-based neural \cite{korse_enhancement_2020} post \Revision{filter has} been used to enhance the speech quality of classical speech codecs. Approaches based on \acp{GAN} have been proposed in \cite{biswas_audio_2020, korse_postgan_2022} and \cite{buthe_nolace_2024}, where the latter one is a low-complex \ac{DNN} model.
%
\subsubsection{Neural Vocoders}
Neural vocoders may be used for decoding the encoded bitstream of a classical codec. Therefore, such approaches can also be made compatible with legacy bitstreams. Starting from \cite{Kleijn2017}, several such approaches have been proposed with increasing quality and lower computational complexity over the recent years \cite{Klejsa2018, garbacea_low_2019, Fejgin2020}. Based on LPCNet \cite{Valin2019}, a neural codec has been presented in \cite{valin_real-time_2019} that can code speech at $1.6\,\mathrm{kbps}$ (kilobits per second) in real time. In \cite{mustafa_streamwise_2021}, SSMGAN has been proposed which was shown to decode the bitstream used in \cite{valin_real-time_2019} at improved quality and in a non-autoregressive manner.
\subsubsection{End-to-end Approaches}
To leverage the full potential of neural coding, end-to-end trained \acp{DNN}, thereby providing both a neural encoder and a neural decoder, have been proposed and now dominate the research on neural audio coding. An early attempt to this was to train an autoencoder \ac{DNN} in a discriminative manner using \ac{MSE} and perceptually motivated losses \cite{kankanahalli_end--end_2021}. To increase perceptual quality, deep generative models have been adopted afterwards as the mainstream training paradigm, where \acp{GAN} represent the most prominent approach in the field of neural audio coding: SoundStream \cite{zeghidour_soundstream_2021} is based on a convolutional encoder-decoder structure and is capable of coding speech at bitrates between $3$ to $12\,\text{kbps}$. SoundStream is also able to perform joint coding and enhancement and may be used for coding general audio \Revision{like music} as well. To improve \Revision{the coding robustness in noisy conditions}, \cite{casebeer_enhancing_2021} proposed to separately \Revision{encode} speaker-related features by temporal averaging. With NESC \cite{pia_nesc_2022}, a low-complex encoder has been combined with SSMGAN \cite{mustafa_streamwise_2021}, providing robust speech coding at low bitrates. EnCodec \cite{defossez_high_2022}, a neural codec for general audio, consists of an encoder-decoder structure that is trained with a single multi-scale spectral discriminator, thereby speeding up the training process. A lightweight transformer-based entropy model may be used for optionally compressing the transmitted data stream even further. A general audio compression model has been proposed in \cite{kumar_high-fidelity_2023} that uses several improvements in training techniques allowing for compression of $44.1\, \mathrm{kHz}$ \Revision{sampled} audio signals to $8\,\mathrm{kbps}$ data streams. Together with an open-source toolbox for neural audio coding, the authors of \cite{du_funcodec_2023} propose a frequency-domain neural speech codec that can be parameterized over a large range of computational complexities. A streamable audio codec working at $48\, \mathrm{kHz}$ has been proposed by \cite{wu_audiodec_2023} together with an efficient training technique. Another neural codec working at very low bitrates ($2\, \mathrm{kbps}$) has been proposed in \cite{chen_tenc_2021}. A neural coding system for real-time communication including packet loss concealment has been developed by \cite{jiang_end--end_2022}. Very recently, APCodec \cite{ai_apcodec_2024} has been proposed. It operates in the frequency domain and encodes the amplitude and phase spectra separately. Also very recently, first neural codecs based on diffusion models have been proposed. Latent diffusion was leveraged in \cite{yang_generative_2023} by employing a diffusion model for dequantization of a latent representation of \Revision{an} end-to-end trained neural codec. In \cite{san_roman_discrete_2023}, the quantized representation of a neural codec has been used for directly generating the decoded waveform with a diffusion model.

\Revision{However, neural speech codecs particularly designed for mobile communications which require low latency and very low complexity for application on mobile devices are underrepresented in the literature. Furthermore, many of the proposed neural network architectures show strong similarities, where the most prominent network type applied for neural coding is SEANet \cite{tagliasacchi_seanet_2020} and is used with small variations in, e.g.,  Soundstream \cite{zeghidour_soundstream_2021}, EnCodec \cite{defossez_high_2022} and AudioDec \cite{wu_audiodec_2023}.}
\subsection{Discrete Representation Learning for Audio Coding}
\Revision{For efficient data transmission, it is often preferable to derive a low-dimensional and discrete signal representation.} While a low-dimensional signal representation can be achieved by appropriate design of the network, i.e., by choosing a low-dimensional bottleneck, learning a discrete representation needs special treatment. \Revision{Discrete representation learning for audio signals is a broad field of research \cite{vq-wav2vec, zhou_comparison_2021, soft_to_hard}, we will therefore focus in the following on approaches used for neural audio coding.}
\subsubsection{Vector Quantization}
The common approach in neural audio coding for discrete representation learning, which has been used by almost all previously mentioned \Revision{works}, is \ac{VQ} \cite{garbacea_low_2019}. Here, each frame-wise output of the neural encoder is approximated by a vector out of a set of candidate vectors, the so-called codebook, which is trained to optimally approximate the multivariate latent signals. To realize \acp{VQ} with deep learning, the \ac{VQ}-\ac{VAE} \cite{oord_neural_2018} has been proposed. The codebook is typically trained by k-means with moving average update and an additional loss ensures compatibility of the encoder with the codebook representation. In order to achieve practical complexity for typical target bitrates, various constraints to the structure of the \ac{VQ} codebook have been introduced. The most prominent approach here is to use \acp{RVQ}, where a stack of \acp{VQ} is trained with each \ac{VQ} module subsequently coding the residual of the preceding module (see Fig.~\ref{fig:rvq} for an illustration). Several other variants of constrained \ac{VQ} have been proposed in the literature for neural audio coding but are not commonly used, e.g., the multi-scale \ac{VQ} \cite{petermann_harp-net_2021}, the cross-scale \ac{VQ} \cite{jiang_cross-scale_2022}\Revision{, and} the grouped \ac{VQ} and beamsearch on the \ac{VQ} codebook \cite{xu_intra-brnn_2024}.

All of the \Revision{aforementioned} \ac{VQ} variants suffer from one or several of the following drawbacks:
\begin{itemize}
	\item \Revision{Inefficient codebook usage}: During training some codevectors tend to be rarely used or neglected completely. To make full use of the codebook, rarely used vectors may be re-initialized during training, which requires additional robust decision rules to achieve \Revision{full} codebook usage.
	\item Hyperparameter choice: For codebook training based on moving average, averaging factors have to be chosen that fit the training speed of the remaining network.
	\item Extra losses: \Revision{Training} a \ac{VQ} \Revision{(typically) requires additional losses}, e.g., a commitment loss or a codebook loss \cite{oord_neural_2018}. 
	\item Loss weighting: These additional losses have to be balanced with other losses used in the training of the neural codec.
	\item Memory requirement: The \ac{VQ} codebook has to be trained and stored in memory.
	\item Codebook search complexity: In order to determine the best codevector for each signal frame, pair-wise distances between all codevectors have to be calculated. 
	\item Gradient flow: For dealing with non-differentiable operations like the codebook search, special mechanisms like the straight-through operator \Revision{\cite{oord_neural_2018}} are needed during training time.
\end{itemize}
\subsubsection{Scalar Quantization}
In this paper, we propose methods based on \ac{SQ}, an approach that has received little attention in \Revision{low-bitrate} neural audio coding. \Revision{However, it is important to mention that \ac{SQ} models recently showed potential in other application domains like image processing \cite{Balle17a,agustsson_universally_2020,balle_nonlinear_2021,mentzer_finite_2023}}. Among the proposed \ac{SQ}-based approaches for neural audio coding there are methods that begin training with a smooth approximation of the non-differentiable quantization operation and change it during training gradually to an operation resembling the quantizer more accurately \cite{kankanahalli_end--end_2021, zhen_scalable_2022}. However, such methods require the design of an annealing schedule that trades off the high variance of the gradients for an approximation close to hard quantization and the inaccurate but smooth approximation at the beginning of training, which \Revision{can be challenging. Several neural audio codecs based on \ac{SQ} have been proposed, however, operating on comparatively high bitrates: The neural codec \cite{byun2023} that is based on a hyperprior model originally proposed in the image processing domain \cite{balle2018variational}. Another neural audio codec based on perceptual losses and \ac{SQ} has been proposed in \cite{shin2022}. In \cite{kim2023} it was proposed to progressively encode/decode the signals at different sampling rates. An \ac{SQ}-based neural speech coding scheme for transmitting redundant information in order to increase robustness against transmission errors has been proposed by \cite{valin2023}.} Recently, a neural audio codec has been proposed that uses \ac{SQ}, but does not mention many details about its realization \cite{liu_high_2023}. \Revision{Many of the above-mentioned neural coding approaches rely on rate-distortion optimization. However, for real-time communications, typically constant packet lengths are required, i.e., constant bitrate models are used here. Hence, variable bitrate models achieved by, e.g., entropy coding, are not considered for such applications. Therefore, there are, especially for applications of \ac{SQ} for neural speech coding at bitrates lower than $5\,\mathrm{kbps}$, no contributions to the best of the authors knowledge.}
\subsection{Paper Outline}
In the following, we list our contributions and give an overview of the remainder of this paper.
\subsubsection{Contributions}
\Revision{Our contributions are related to efficient solutions for speech communication. In particular, we propose simple and efficient quantization techniques and a computationally low-complex neural speech codec operating at low bitrates.  More specifically, our contributions include the following:}
\begin{itemize}
	\item We propose two simple and efficient realizations of \Revision{\ac{SQ}} for end-to-end discrete representation learning, \Revision{and we show their effectiveness} for neural speech coding. The proposed methods are computationally efficient and can simply be used as a plain neural network block without extra components in the training pipeline like additional losses or schedulers.
	\item A discussion of conceptual differences between commonly used \ac{VQ} approaches and the proposed methods are provided illustrating the benefits of the developed quantizer. Thereby, we advertise the use of \ac{SQ} for neural audio coding \Revision{at low bitrates}, which did not attract much attention by the community so far.
	\item Theoretical analysis and interpretations are provided to highlight relations to \Revision{\ac{VQ}-\acp{VAE}}, between the different proposed \ac{SQ} variants and the influence of the proposed quantizer on the training dynamics.
	\item A new causal architecture for neural speech coding \Revision{for real-time speech communication applications} is proposed. \Revision{The proposed network operates in \ac{STFT} domain in order to emphasize harmonic structure of the signals and represents a pure down/upsampling model which allows for very efficient realizations.}
	\item We give an overview about different quantization techniques for neural audio coding and present an experimental evaluation of them.
\end{itemize}
\subsubsection{Overview}
\ac{VQ}-based methods for neural audio coding representing the state of the art \Revision{at low bitrates} are reviewed and discussed in Sec.~\ref{sec:vq}. In Sec.~\ref{sec:sq}, the proposed \ac{SQ}-based approaches are introduced and analyzed. The proposed neural speech coding network architecture is presented in Sec.~\ref{sec:network_architecture}. Experimental results are compiled in Sec.~\ref{sec:experiments} and a conclusion is given in Sec.~\ref{sec:conclusion}.
\section{Conventional Quantization with VQ}
\label{sec:vq}
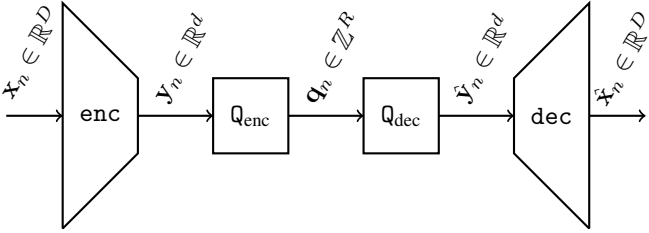
\begin{figure}
\centering
\begin{tikzpicture}
\draw[thick] (0,0)--(0,3)--(1,2)--(1,1)--(0,0);
\draw[thick] (2,1)rectangle(3,2);
\draw[thick] (4,1)rectangle(5,2);
\draw[thick] (7,0)--(7,3)--(6,2)--(6,1)--(7,0);

\node at(0.5,1.5) {$\Encoder$};
\node at(2.5,1.5) {$\Qenc$};
\node at(4.5,1.5) {$\Qdec$};
\node at(6.5,1.5) {$\Decoder$};

\draw[->, thick] (-0.75,1.5)--node[above, rotate=60, xshift = 20pt, yshift=5pt]{$\XVec_\FrameIdx \in \mathbb{R}^{D}$}(0,1.5);
\draw[->, thick] (1,1.5)--node[above, rotate=60, xshift = 20pt, yshift=0pt]{$\LatVec_\FrameIdx \in \mathbb{R}^{d}$}(2,1.5);
\draw[->, thick] (3,1.5)--node[above, rotate=60, xshift = 20pt, yshift=0pt]{$\CdbkIdxVec_\FrameIdx \in \mathbb{Z}^{\NumQuants}$}(4,1.5);
\draw[->, thick] (5,1.5)--node[above, rotate=60, xshift = 20pt, yshift=0pt]{$\LatVecQ_\FrameIdx \in \mathbb{R}^{d}$}(6,1.5);
\draw[->, thick] (7,1.5)--node[above, rotate=60, xshift = 20pt, yshift=0pt]{$\XVecRec_\FrameIdx \in \mathbb{R}^{D}$}(7.75,1.5);
\end{tikzpicture}
\caption{Generic overview of a neural audio codec: The input signal frame $\XVec_\FrameIdx$ is processed by a neural encoder $\Encoder$ yielding a latent representation $\LatVec_\FrameIdx$. This latent representation is mapped to a discrete representation $\CdbkIdxVec_\FrameIdx$ which is used for transmission. The decoder part receives the discrete representation and dequantizes it by $\Qdec$. The result is fed to a neural decoder $\Decoder$ which reconstructs the input signal frame.}
\label{fig:codec_block_diagram}
\end{figure}
Typically, a neural audio codec (see Fig.~\ref{fig:codec_block_diagram} for an illustration) consists of a neural encoder $\Encoder$ that maps each frame of the input signal waveform $\XVec_\FrameIdx\in\mathbb{R}^\DimData$ to a (continuous) latent representation $\LatVec_\FrameIdx$
\begin{equation}
\LatVec_\FrameIdx = \Encoder(\XVec_\FrameIdx) \in \mathbb{R}^\DimLatent.
\end{equation}
Here, $\FrameIdx$ denotes the frame index and $\DimLatent$ and $\DimData$ the dimensions of the latent signal frame and the input dimension, respectively. The latent representation $\LatVec_\FrameIdx\in \mathbb{R}^\DimLatent$ is quantized yielding $\LatVecQ_\FrameIdx\in \mathbb{R}^\DimLatent$, where in an intermediate step a vector of codebook indices $\CdbkIdxVec_\FrameIdx\in \mathbb{Z}^\NumQuants$ is calculated which is used for transmission in a bitstream representation
\begin{equation}
\CdbkIdxVec_\FrameIdx = \Qenc(\LatVec_\FrameIdx)\in \mathbb{Z}^\NumQuants \qquad \LatVecQ_\FrameIdx = \Qdec(\CdbkIdxVec_\FrameIdx)\in \mathbb{R}^\DimLatent.
\end{equation}
Hence, the concatenation $\Qdec(\Qenc(\cdot))$ represents the quantizer and $\NumQuants$ the number of indices used to characterize one signal frame. The dequantized latent representation $\LatVecQ_\FrameIdx$ is then decoded by a neural decoder in order to approximate the input waveform
\begin{equation}
\XVec_\FrameIdx \approx \XVecRec_\FrameIdx = \Decoder(\LatVecQ_\FrameIdx)\in\mathbb{R}^\DimData.
\end{equation}
In \ac{VQ}, the latent representation $\LatVec_\FrameIdx$ is represented by the best candidate out of the set $\{\LatVecQ^1, \dots, \LatVecQ^\NumLevels\}$ of $\NumLevels$ candidate codevectors, i.e., the codebook
\begin{align}
&\Qenc^\text{VQ}: \mathbb{R}^\DimLatent\rightarrow \SetIndVQ\\
&\Qdec^\text{VQ}: \SetIndVQ \rightarrow \left\{\LatVecQ^1,\dots, \LatVecQ^\NumLevels\right\}\nonumber.
\end{align}
Here, $\Qenc^\text{VQ}$ is typically realized by outputting the index of the codevector $\CdbkIdx^\ast\in\SetIndVQ=\{1,\dots,\NumLevels\}$ that corresponds to the minimal pair-wise distance between $\LatVec_\FrameIdx$ and all codevectors. $\Qdec^\text{VQ}$ maps \Revision{the} codebook index to the corresponding codevector. The concatenation $\mathtt{VQ}(\cdot) = \Qdec^\text{VQ}(\Qenc^\text{VQ}(\cdot))$ represents the \ac{VQ} module.

\begin{figure}
\begin{tikzpicture}
	\node[draw, thick](vq1) at (0,0) {$\text{VQ}_1$};
	\node[draw, thick](vq2) at (2.8,0) {$\text{VQ}_2$};
	\node[draw, thick](vq3) at (5.6,0) {$\text{VQ}_3$};
	
	\draw[->, thick] (-1.0,0)--(vq1);
	\draw[->, thick] (-0.75,0)--(-0.75,-0.75)--(1.2,-0.75)--(1.2,-0.2);
	\draw[->, thick] (vq1)--(vq2);
	\draw[->, thick] (1.8,0)--(1.8,-0.75)--(4,-0.75)--(4,-0.2);
	\draw[->, thick] (vq2)--(vq3);
	\draw[->, thick] (4.75,0)--(4.75,-0.75)--(6.75,-0.75)--(6.75,-0.2);
	\draw[->, thick] (vq3)--(7.25,0)node[right]{...};
	
	\draw[thick, fill=black] (-0.75,0) circle(1pt);
	\draw[thick, fill=black] (0.75,0) circle(1pt);
	\draw[thick, fill=white] (1.2,0) circle(5pt) node{$-$};
	\draw[thick, fill=black] (1.8,0) circle(1pt);
	\draw[thick, fill=black] (3.5,0) circle(1pt);
	\draw[thick, fill=white] (4,0) circle(5pt) node{$-$};
	\draw[thick, fill=black] (4.75,0) circle(1pt);
	\draw[thick, fill=white] (6.75,0) circle(5pt) node{$-$};
	\draw[thick, fill=black] (6.25,0) circle(1pt);
	
	\draw[thick, fill=white] (3.5,0.75) circle(5pt) node{$+$};
	\draw[thick, fill=white] (6.25,0.75) circle(5pt) node{$+$};
	\draw[->, thick] (0.75,0)--(0.75,0.75)--(3.3,0.75);
	\draw[->, thick] (3.5,0)--(3.5,0.55);
	\draw[->, thick] (6.25,0)--(6.25,0.55);
	\draw[->, thick] (3.75,0.75)--(6.05,0.75);
	\draw[->, thick] (6.5,0.75)--(7.25,0.75)node[above]{output};
\end{tikzpicture}
\caption{Illustration of an \ac{RVQ}.}
\label{fig:rvq}
\end{figure}
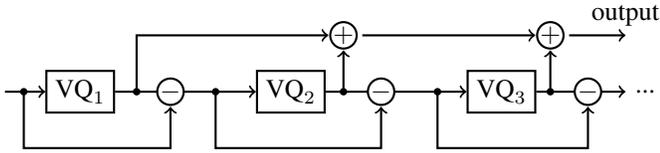

However, to realize \acp{VQ} providing practical codebook sizes under limited memory requirements usually requires to add more structure to the \ac{VQ}, e.g., to realize a bitstream of $1.5\,\mathrm{kbps}$ with $20\,\mathrm{ms}$ framing would require $2^{30}$ codebook vectors in a single \ac{VQ} codebook\Revision{, which is practically impossible to use in an exhaustive search}. The most prominent solution to this problem is \ac{RVQ} (see Fig.~\ref{fig:rvq}) in neural audio coding (as well as for classical audio coding). Here, a cascade of \ac{VQ} \Revision{$\NumQuants$ stages (i.e., $\NumQuants=1$ for plain \ac{VQ})} is used to encode the latent representation, where each approximates a residual representation w.r.t. the input of the previous module. This also allows for bitrate scalability by dropping out \ac{VQ} modules. Especially in classical speech coding other constraints were considered, like split \ac{VQ}, limiting the vector dimension by splitting into sub-vectors, tree structured codebooks or lattice \ac{VQ}~\cite{gersho_vq_1991}. 

The selection of the optimum codevectors, i.e., $\Qenc^\text{VQ}$, is not a differentiable operation. Hence, in order to make the \ac{VQ} module trainable end-to-end, specialized training techniques have to be used. Here, the straight-through (ST) gradient operator is typically used \cite{oord_neural_2018}. \Revision{This operator basically approximates a non-differentiable network component as an identity for the calculation of the gradients.} However, due to the ST operator the \ac{VQ} codebook does not receive any gradients and would not be updated by gradient descent steps. To solve this issue, the training losses of the neural codec \Revision{are amended with the additional loss terms}
\begin{equation}
\mathcal{L}_{\text{VQ}} = \underbrace{\left\Vert \text{sg}\{\LatVec_\FrameIdx\} - \LatVecQ^\ast_\FrameIdx \right\Vert_2^2}_{\text{VQ loss}} + \beta\underbrace{\left\Vert \LatVec_\FrameIdx - \text{sg}\{\LatVecQ^\ast_\FrameIdx\} \right\Vert_2^2}_\text{commitment loss}.
\end{equation}
Here, $\text{sg}\{\cdot\}$ denotes the stop gradient operator and $\beta\geq 0$ is a weighting factor. The first term is used for fitting the \ac{VQ} codebook to the latent representations and the second term enforces that the encoder commits to the codebook. Intuitively, the first term moves the codevectors towards the latent representations and the second term moves the latent representation towards the codevectors. As there is no restriction to the value range of the encoder outputs or codevectors, the commitment loss is particularly important to avoid arbitrary growth of the encoder outputs. In practical implementations like \cite{oord_neural_2018}, the \ac{VQ} loss is often dropped and the codebook is learned by k-means with exponential moving average.
%
\section{Proposed SQ-Based Quantization Technique}
\label{sec:sq}
%
%
In this paper, we propose a simple and efficient way to realize $\Qenc$ and $\Qdec$ with \ac{SQ}. To this end, we project the latent representation $\LatVec_\FrameIdx\in\mathbb{R}^\DimLatent$ to a lower dimensional representation \Revision{$\SQVec_\FrameIdx=\mathtt{proj}(\LatVec_\FrameIdx)\in\mathbb{R}^\NumQuants$} (this has been done similarly in the image processing domain \cite{mentzer_finite_2023}) and restrict the value range by the application of a suitable activation function, e.g., $\mathtt{tanh}$. The projection $\mathtt{proj}$ enables the use of \ac{SQ} techniques also for low bitrates and can be realized in principle by any neural network, e.g., a linear projection in the simplest case. The \Revision{projected latent representation} $\SQVec_\FrameIdx\in\mathbb{R}^\NumQuants$, $\NumQuants\leq\DimLatent$ is then quantized element-wise, i.e., we map each element $\SQVec_{\QuantIdx,\FrameIdx}$ of $\SQVec_\FrameIdx$ to an index from $\SetIndSQ = \left\{\NumLevels_\text{min},\dots, \NumLevels_\text{max}\right\}$ and then map those indices back to a quantized signal representation $\SQVecQ_{\QuantIdx,\FrameIdx}$ in a second step. For simplicity, we choose the quantization levels in this paper to be fixed, uniform and identical for each \Revision{of the $\NumQuants$ latent dimensions}, which is not strictly necessary 
\begin{align}
	\mathtt{Quant}:\ &[-1,1]\rightarrow \SetIndSQ\ \text{with}\ \CdbkIdx_{\FrameIdx} = \left\lfloor\frac{\SQVar_{\QuantIdx,\FrameIdx}}{\Delta}\right\rfloor + \frac{1}{2} \label{eq:SQ_quant}\\ 
	\mathtt{Dequant}:\ &\SetIndSQ\rightarrow \left\lbrace\CdbkIdx\QuantLevel\vert \CdbkIdx\in \SetIndSQ\right\rbrace\ \text{with}\ \SQVarQ_{\QuantIdx,\FrameIdx} = \CdbkIdx_{\QuantIdx,\FrameIdx}\QuantLevel \nonumber.
\end{align}
Here, $\NumLevels_\text{min}$ and $\NumLevels_\text{max}$ represent the smallest and largest index, respectively, and $\QuantLevel$ the quantization step size. Note that we show here exemplary a mid-rise quantizer \Revision{(see \eqref{eq:SQ_quant})} but any other form of scalar quantization could be used as well in principle. The resulting quantized representation $\SQVecQ_{\QuantIdx,\FrameIdx}$ is then mapped by a neural network, typically symmetric in architecture to the input projection to $\LatVecQ_\FrameIdx\in\mathbb{R}^\DimLatent$. For notational convenience, we will denote \ac{SQ} by $\mathtt{SQ}(\cdot) = \mathtt{Dequant}(\mathtt{Quant}(\cdot))$ and the full module by $\mathtt{PSQ}(\cdot) = \Qdec^\text{SQ}\left(\Qenc^\text{SQ}(\cdot)\right)$ (Projected \ac{SQ}). Here, $\Qenc^\text{SQ}$ consists of the input projection, $\mathtt{tanh}$ and element-wise $\mathtt{Quant}$ and $\Qdec^\text{SQ}$ contains element-wise $\mathtt{Dequant}$ and an output projection. \Revision{Please note that the discrete representation learned in this way does not have a direct correspondence to the inputs waveform, e.g., input signal values small in magnitude are not systematically assigned to dequantized values of small magnitude.}
\subsection{SQ Realizations}
\label{sec:sq_realizations}
Similar as for \ac{VQ}, the operation $\mathtt{Quant}$ is not differentiable and special care has to be dedicated to a trainable realization of $\mathtt{SQ}$. Soft-to-hard annealing \cite{kankanahalli_end--end_2021, zhen_scalable_2022} has been used as a remedy for the non-differentiability issue. However, the choice of an effective annealing schedule is difficult. Therefore, we will discuss two simple alternative methods for realizing $\mathtt{SQ}$ in the following that do not require any additional loss functions, schedules etc. and can, hence, be used as a plug-in layer during training, which simplifies the training of neural audio codecs significantly.
\subsubsection{Straight-through (ST) Gradient}
Similarly to \ac{VQ}-\ac{VAE}, a ST gradient operator can be used to deal with the non-differentiable \ac{SQ} operation \eqref{eq:SQ_quant}. An advantage of this procedure is that the quantization that is applied during inference is exactly the same also seen during training. However, the effect of the quantizer is only seen in the forward pass. 
\subsubsection{Pseudo Quantization Noise Training}
An alternative approach is to mimic the quantizer by adding uniform noise to the latent representation during training (see \cite{balle_nonlinear_2021} for an application of a related technique from the image coding domain)
\begin{equation}
	\SQVecQ = \SQVec + \NoiseVec\quad \text{with}\quad \NoiseVec\sim\mathcal{U}\left\lbrace\bigg[-\frac{\QuantLevel}{2},\frac{\QuantLevel}{2}\bigg)^\NumQuants\right\rbrace.
	\label{eq:SQ_noise_addition}
\end{equation}
Approximating quantization noise with additive uniform noise is often done in the literature for the sake of statistical analysis and is well justified if the distance between the quantization levels is small \cite{rabiner_theory_2011} (see also Sec.~\ref{sec:SQ_dithered}). During \Revision{inference} regular quantization is applied to the latent representation in order to produce a discrete signal representation for transmission.
\subsection{Interpretations and Relations}
In the following we interpret and analyze the proposed \ac{SQ} realizations.
\subsubsection{VQ and Transform Coding}
All possible combinations of chosen quantization levels of an \ac{SQ} realize a codebook similar to the one of a \ac{VQ}
\begin{equation}
	\left\{\SQVecQ^1,\dots, \SQVecQ^\NumLevels\right\} = \left\{\NumLevels_\text{min}\QuantLevel,\dots, \NumLevels_\text{max}\QuantLevel\right\}^\NumQuants.
\end{equation}
Here, \Revision{$\NumLevels = (\NumLevels_\text{min} + \NumLevels_\text{max} +1)^\NumQuants$}.
However, for \ac{SQ} the optimal codevector is determined by choosing the best quantization levels element-wise, which then results also in the optimum \ac{SQ}-based codevector. 
The resulting \ac{SQ} codebook of the fixed, uniform and dimension-wise identical quantization levels represents a uniform grid on a hypercube, i.e., $[-1,1]^\NumQuants$ in the latent of $\mathtt{PSQ}$. The decoder $\Qdec$ of the $\mathtt{PSQ}$ transforms these vectors back to a $d$-dimensional representation. In the most simple case of a linear transform for the decoder, i.e., a matrix multiplication with $\mathbf{D}\in\mathbb{R}^{\DimLatent\times\NumQuants}$, the effective \ac{SQ} codebook equivalent is a set of vectors in the column space of $\mathbf{D}$ (note that the decoder may also be a \ac{DNN} allowing for complex nonlinear transformations). Hence, the codevectors of an $\mathtt{PSQ}$ may be structurally related to each other. On the other side, a \ac{VQ} codebook consists of vectors that are \Revision{not directly structurally related with each other}. To add flexibility to the \ac{SQ} codebook, the quantization levels of the \ac{SQ} could be trained, could be chosen non-uniform or differently for each latent dimension. However, none of these adjustments lead to significant benefits over the described approach and the reduced flexibility of the effective $\mathtt{PSQ}$ codebook did not show a detrimental effect relative to a \ac{VQ} in our experiments.
\begin{figure}
	\centering
	\includegraphics[scale=0.175]{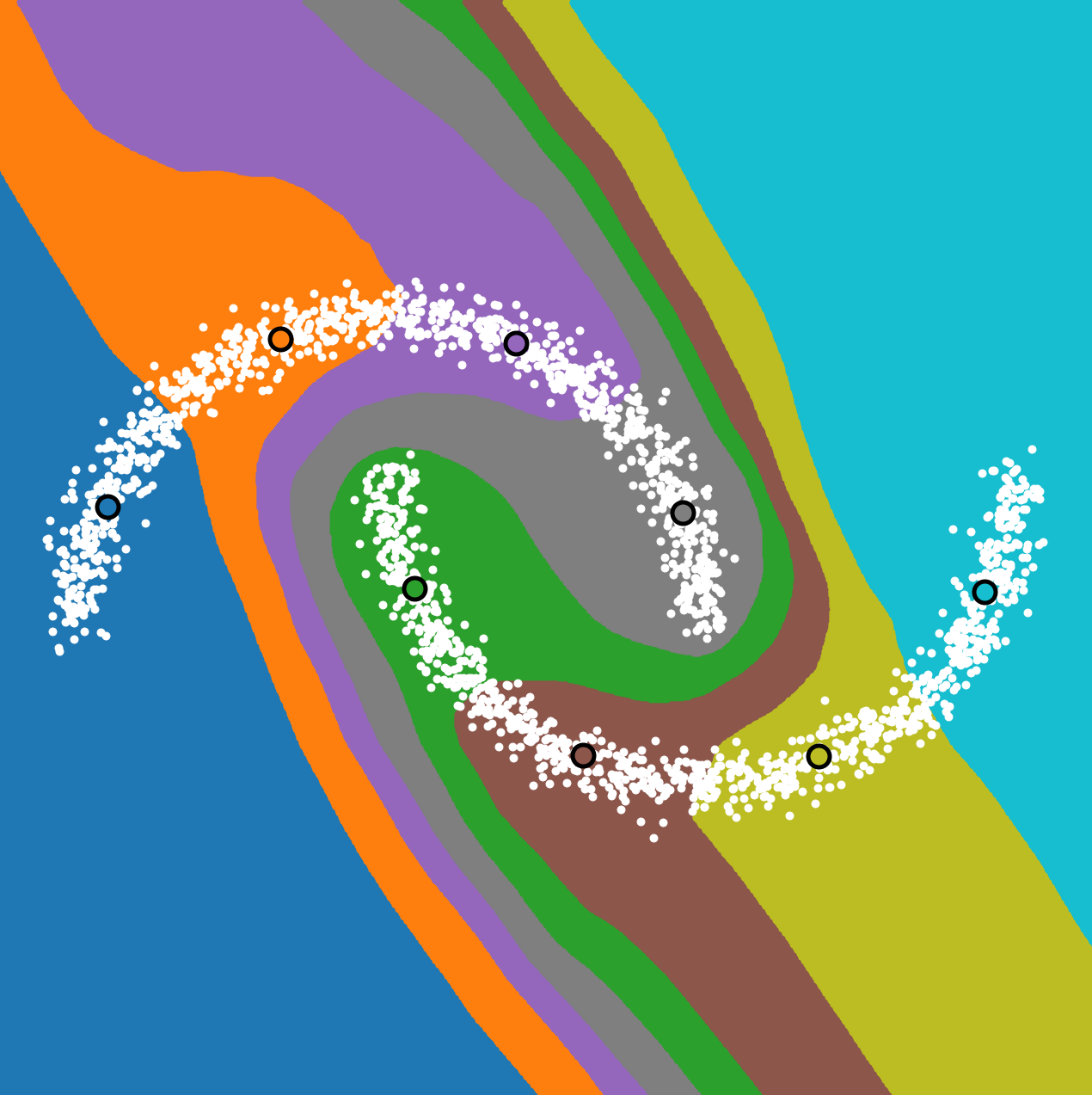}
	\caption{Quantization of a toy data set (shown in white) with the proposed \ac{SQ}-based method using three bits. The decision regions corresponding to each codebook vector is shown with a different color.}
	\label{fig:voronoi}
\end{figure}

For classical audio coding, \ac{VQ} was shown to be beneficial especially for low bitrates \Revision{where the} following three benefits are attributed to \ac{VQ} relative to \ac{SQ} \cite{lookabaugh_vq_1989}: The better packing of high-dimensional spaces by the Voronoi cells of a VQ, i.e., the space-filling advantage, the increased flexibility of approximating \acp{PDF} of arbitrary shapes, i.e., the shape advantage, and the capability of modeling \Revision{correlated samples}, i.e., the memory advantage. However, it should be noted that the $\mathtt{PSQ}$ consists of the quantizer itself, but also of an input and an output projection that may realize nonlinear transformations, i.e., the proposed method may be interpreted as \textit{nonlinear transform coding} \cite{balle_nonlinear_2021}. 
As the encoder/decoder of the neural audio codec as well as the input/output transform of the $\mathtt{PSQ}$ may model arbitrary nonlinear functions with temporal context, $\mathtt{PSQ}$ shows the shape and the memory advantage attributed to \acp{VQ}. \ac{SQ} as we defined it relies on a regular grid of codebook vectors and, hence, would not directly show the space-filling advantage as \acp{VQ}. However, as the input and output projection of the $\mathtt{PSQ}$ may shape the \ac{SQ} latent space such that signal reconstruction is optimal, $\mathtt{PSQ}$ is expected to show also space-filling behavior: The regular grid of codebook vectors is mapped by the output projection to an arbitrarily distribution of effective codebook vectors. To illustrate this, we trained a simplistic $\mathtt{PSQ}$ to quantize on $3\,\mathrm{bits}$, two-dimensional toy data, where the data distribution is shaped in `two moons' (see Fig~\ref{fig:voronoi}). The $\mathtt{PSQ}$ is clearly able to adjust to the shape of the data distribution and models correlations within the data. The color-coded decision regions show that the learned input and output transforms are indeed highly nonlinear. While allowing for precise quantization of data following the trained distribution by \ac{SQ}, the decision boundaries may be ill-conditioned for samples far away from the trained distribution. However, this effect depends on the learned mapping, the trained distribution and the application scenario. Note that also the behavior of traditional \acp{VQ} is not controlled for out-of-domain areas, i.e., a similar behavior can be observed here \Revision{and is known as the saturation regime}.
%
%
\subsubsection{Dithered Quantization}
\label{sec:SQ_dithered}
When training with the noise addition surrogate \eqref{eq:SQ_noise_addition}, two statistically independent random vectors are added. Hence, the resulting \ac{PDF} of this sum $\SQVecQ$ is the convolution of the \acp{PDF} of latent $\SQVecQ$ and uniformly distributed random vector $\NoiseVec$
\begin{align}
	p_{\SQVecQ}(\SQVecQ) &= (p_{\SQVec} \ast p_{\NoiseVec})(\SQVecQ) = \int_{-\infty}^{\infty} p_{\SQVec}(\SQVec)  p_{\NoiseVec}(\SQVecQ-\SQVec)\text{d}\SQVec\\
	&= \frac{1}{\QuantLevel} \int_{\SQVecQ -\frac{\QuantLevel}{2}\mathbf{1}}^{\SQVecQ +\frac{\QuantLevel}{2}\mathbf{1}} p(\SQVec) \text{d}\SQVec.
\end{align}
Here, $\mathbf{1}$ denotes an all-one vector of appropriate dimension. In the following, we show that mimicking scalar quantization with uniform noise addition is equivalent to dithered quantization \cite{gray_dithered_1993}, i.e., adding uniform noise prior to quantization and subtracting the same noise realization again afterwards
\begin{equation}
	\hat{\tilde{\SQVec}}^{\text{D}} = \underbrace{\mathtt{SQ}(\SQVec + \tilde{\NoiseVec})}_{\tilde{\SQVec}^\text{D}} - \tilde{\NoiseVec}\quad \text{with}\quad \tilde{\NoiseVec}\sim\mathcal{U}\left\lbrace\bigg[-\frac{\QuantLevel}{2},\frac{\QuantLevel}{2}\bigg)^\NumQuants\right\rbrace.
	\label{eq:dithered_SQ}
\end{equation}
Note that in a practical realization of dithered quantization, the noise signal does not have to be transmitted but by using the same seed on transmitter and receiver side, the same pseudo-random noise sequence can be generated for the dithered quantizer. 

For simplicity, we first derive the conditional probability, denoted by $P\left(\hat{\tilde{\SQVar}}^{\text{D}}\vert \tilde{\NoiseVar}\right)$, of observing the quantization result $\hat{\tilde{\SQVar}}^{\text{D}}$ given a realization of uniform noise $\tilde{\NoiseVar}$
\begin{align}
	P\left(\hat{\tilde{\SQVar}}^{\text{D}}\vert \tilde{\NoiseVar}\right) 
	&=  P\left\lbrace\tilde{\SQVar}^\text{D} -\frac{\QuantLevel}{2} \leq \SQVar + \tilde{\NoiseVar} < \tilde{\SQVar}^\text{D} +\frac{\QuantLevel}{2}\right\rbrace.
\end{align}
This is the probability that a latent variable $\SQVar$ falls inside of the quantization interval $[\hat{\tilde{\SQVar}}^\text{D} -\frac{\QuantLevel}{2}, \hat{\tilde{\SQVar}}^\text{D} +\frac{\QuantLevel}{2})$, while fixing the noise~$\tilde{\NoiseVar}$
\begin{align}
	P\left(\hat{\tilde{\SQVar}}^{\text{D}}\vert \tilde{\NoiseVar}\right) 
	&=  P\left\lbrace\hat{\tilde{\SQVar}}^\text{D} -\frac{\QuantLevel}{2} \leq \SQVar < \hat{\tilde{\SQVar}}^\text{D} +\frac{\QuantLevel}{2}\right\rbrace\\
	&=  \int_{\hat{\tilde{\SQVar}}^\text{D} -\frac{\QuantLevel}{2}}^{\hat{\tilde{\SQVar}}^\text{D} +\frac{\QuantLevel}{2}} p(\SQVar) \text{d}\SQVar.
\end{align}
Hence, the resulting probability is obtained by integrating the \ac{PDF} of the latent $\SQVar$ over the quantization interval.

This result is straightforwardly extended to the vector-valued case
\begin{equation}
	P\left(\hat{\tilde{\SQVec}}^{\text{D}}\vert \tilde{\NoiseVec}\right) = \int_{\hat{\tilde{\SQVec}}^\text{D} -\frac{\QuantLevel}{2}\mathbf{1}}^{\hat{\tilde{\SQVec}}^\text{D} +\frac{\QuantLevel}{2}\mathbf{1}} p(\SQVec) \text{d}\SQVec.
\end{equation}
We observe that
\begin{equation}
	P\left(\hat{\tilde{\SQVec}}^{\text{D}}\big\vert \tilde{\NoiseVec}\right) = \QuantLevel\ p_{\SQVecQ}\left(\hat{\tilde{\SQVec}}^{\text{D}}\right),
\end{equation}
i.e., the conditional probability given the particular noise realization $\tilde{\NoiseVec}$ is proportional to the \ac{PDF} of the noise surrogate model. 
Hence, dithered quantization and noise addition follow the same distribution, i.e., are equivalent in this sense.  
\subsubsection{VAE}
In the following section, we give a \ac{VAE} interpretation of the proposed quantizers similar to \ac{VQ}-\ac{VAE} \cite{oord_neural_2018}. We begin by the generic ELBO loss of a \ac{VAE} \cite{kingma_auto-encoding_2022} 
\begin{gather}
	\mathcal{L}_{\text{ELBO}} = \EOP{p(\XVec)}{\, \KLDOP{q(\SQVecQ\vert\XVec)}{p(\SQVecQ)} - \EOP{q(\SQVecQ\vert\XVec)}{\log p(\XVec\vert\SQVecQ)}\, },
\label{eq:ELBO_loss}
\raisetag{-5pt}
\end{gather}
where the first term is a prior matching term and the second term is a reconstruction loss dependent on the form of the chosen \ac{PDF}, e.g., MSE if $p(\XVec\vert\SQVecQ)$ is Gaussian. In practice, the reconstruction term may be replaced by sophisticated training losses, see Sec.~\ref{sec:losses}. In the following, we will focus on the prior matching term and begin with an interpretation of the noise addition surrogate quantizer. 
\paragraph{Pseudo Quantization Noise Training} Here, the posterior density is uniformly distributed around the latent $\SQVec$ (which is a function of $\XVec$)
\begin{equation}
	q(\SQVecQ\vert\XVec) = \mathcal{U}\left\lbrace\bigg[\SQVec - \frac{\QuantLevel}{2}\mathbf{1}, \SQVec + \frac{\QuantLevel}{2}\mathbf{1}\bigg)\right\rbrace.
\end{equation}
During training we use a uniform distribution over the hypercube containing all possible latent vectors of the $\mathtt{PSQ}$ (ensured by the $\mathtt{tanh}$ activation) 
\begin{equation}
	p(\SQVecQ) = \mathcal{U}\left\lbrace\left[-1,1\right]^{\NumQuants}\right\rbrace.
\end{equation}
This uniform prior enforces the full usage of the hypercube defining the latent space. Based on these assumptions, we compute the expression of the corresponding prior matching loss
\begin{align}
	\KLDOP{q(\SQVecQ\vert\XVec)}{p(\SQVecQ)} &= \int_{[-1,1]^\NumQuants} q(\SQVecQ\vert\XVec) \log \frac{q(\SQVecQ\vert\XVec)}{p(\SQVecQ)} \text{d}\SQVecQ\\ 
	& = \left(\frac{2}{\QuantLevel}\right)^\NumQuants \log \left(\frac{2}{\QuantLevel}\right)^\NumQuants= \text{const.},
\end{align}
which is constant, i.e., the \ac{VAE} loss function \eqref{eq:ELBO_loss} effectively reduces to the reconstruction loss.
\paragraph{ST Gradient} Similarly, we investigate the \ac{SQ} realization by ST gradient and choose the following probability mass function (note that $\CdbkIdxVec$ is discrete) 
\begin{equation}
	q(\CdbkIdxVec\vert\XVec) = \left\lbrace\begin{matrix}
		1 \quad \text{if}\ \CdbkIdxVec=\CdbkIdxVec^\ast=\Qenc(\Encoder(\XVec))\\
		0 \quad \text{else}
	\end{matrix} \right. .
\end{equation}
We choose again a uniform prior (now over all possible discrete \ac{SQ} outputs) to enforce the comprehensive usage of all possible \ac{SQ} outputs for the latent representation in order to make efficient use of the available bitrate
\begin{equation}
	p(\CdbkIdxVec) = \mathcal{U}\left\lbrace \{\NumLevels_\text{min},\dots, \NumLevels_\text{max}\}^\NumQuants \right\rbrace.
\end{equation}
With this, we obtain again a constant prior matching loss contribution
\begin{align}
	\KLDOP{q(\CdbkIdxVec\vert\XVec)}{p(\CdbkIdxVec)} &= \sum_\CdbkIdxVec q(\CdbkIdxVec\vert\XVec) \log \frac{q(\CdbkIdxVec\vert\XVec)}{p(\CdbkIdxVec)}\\
	&= -\log p\left(\CdbkIdxVec^\ast\right) = \text{const.},
\end{align}
i.e., the \ac{VAE} loss function \eqref{eq:ELBO_loss} reduces again to a plain reconstruction loss. Note that in both cases the choice of a uniform prior was made possible due to the restriction of the value range of the latent vectors.
\subsubsection{Tikhonov Regularization}
Finally, we show that the proposed quantizers may have a regularizing effect on the decoder. For tractability, we assume a purely linear autoencoder consisting of an encoder matrix $\mathbf{E}\in\mathbb{R}^{\NumQuants\times d}$ and a decoder matrix $\mathbf{D}\in\mathbb{R}^{d \times \NumQuants}$. The effect of the quantizer is modeled by additive uniform noise $\NoiseVec$ with identical element-wise variance $\Var$, which is uncorrelated to the latent representation. For simplicity, we use an \ac{MSE} loss for the linear autoencoder and obtain
\begin{align}
	&\EOP{p(\SQVec), p(\NoiseVec)}{\Vert \mathbf{D}\left(\mathbf{E}\SQVec + \NoiseVec\right) - \SQVec\Vert_2^2}\\
	&= \EOP{p(\SQVec)}{\Vert \mathbf{D}\mathbf{E}\SQVec - \SQVec\Vert_2^2} + \text{tr}\left\lbrace\mathbf{D}\transp\mathbf{D}\EOP{p(\NoiseVec)}{\NoiseVec \NoiseVec\transp}\right\rbrace \label{eq:regularization_2nd}\\
	&=\EOP{p(\SQVec)}{\Vert \mathbf{D}\mathbf{E}\SQVec - \SQVec\Vert_2^2} + \Var \Vert\mathbf{D}\Vert_{\text{F}}^2. \label{eq:regularization_3rd}
\end{align}
Here, $\Vert \cdot \Vert_\text{F}$ denotes the Frobenius norm and $\text{tr}\{\cdot\}$ the trace. Note that \eqref{eq:regularization_2nd} follows from the fact that the quantization noise and the latent are uncorrelated and \eqref{eq:regularization_3rd} given by the assumption of the isotropic variance of the additive noise. The first term of \eqref{eq:regularization_3rd} can be identified as an \ac{MSE} reconstruction loss of the linear autoencoder, while the second term realizes a Tikhonov regularizer on the decoder matrix $\mathbf{D}$. The linear model used for this discussion may be simplistic but still gives some intuition on the behavior of the much more complex autoencoder models used in neural audio coding or on simple input and output projections of the $\mathtt{PSQ}$ module.
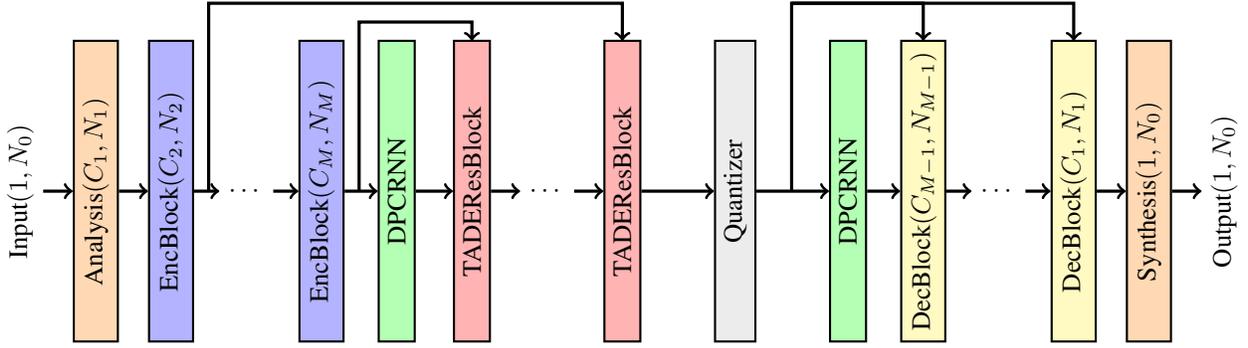
\begin{figure*}
	\centering
	\begin{tikzpicture}
	\node[rotate=90](input) at (0,0){$\text{Input}(1,\NumFrames_0)$};
	\node[draw, thick, rotate=90, minimum width=4cm, below of = input, fill=orange!30!white](l1){$\text{Analysis}(\NumCh_1,\NumFrames_1)$};
	\node[draw, thick, rotate=90, minimum width=4cm, below of = l1, fill=blue!30!white](l2){$\text{EncBlock}(\NumCh_2,\NumFrames_2)$};
	\node[right of = l2](l3){$\cdots$};
	\node[draw, thick, rotate=90, minimum width=4cm, below of = l3, fill=blue!30!white](l4){$\text{EncBlock}(\NumCh_\NumBlocks,\NumFrames_\NumBlocks)$};
	\node[draw, thick, rotate=90, minimum width=4cm, below of = l4, fill=green!30!white](l5){$\text{DPCRNN}$};
	\node[draw, thick, rotate=90, minimum width=4cm, below of = l5, fill=red!30!white](l6){$\text{TADEResBlock}$};
	\node[right of = l6](l7){$\cdots$};
	\node[draw, thick, rotate=90, minimum width=4cm, below of = l7, fill=red!30!white](l8){$\text{TADEResBlock}$};
	\node[draw, thick, rotate=90, minimum width=4cm, below of = l8, yshift=-0.5cm, fill=lightgray!30!white](l9){$\text{Quantizer}$};
	\node[draw, thick, rotate=90, minimum width=4cm, below of = l9, yshift=-0.5cm, fill=green!30!white](l10){$\text{DPCRNN}$};
	\node[draw, thick, rotate=90, minimum width=4cm, below of = l10, fill=yellow!30!white](l11){$\text{DecBlock}(\NumCh_{\NumBlocks-1},\NumFrames_{\NumBlocks-1})$};
	\node[right of = l11](l12){$\cdots$};
	\node[draw, thick, rotate=90, minimum width=4cm, below of = l12, fill=yellow!30!white](l13){$\text{DecBlock}(\NumCh_1,\NumFrames_1)$};
	\node[draw, thick, rotate=90, minimum width=4cm, below of = l13, fill=orange!30!white](l14){$\text{Synthesis}(1,\NumFrames_0)$};
	\node[rotate=90, below of = l14](output){$\text{Output}(1,\NumFrames_0)$};
	
	\draw[->, very thick] (input)--(l1);
	\draw[->, very thick] (l1)--(l2);
	\draw[->, very thick] (l2)--(l3);
	\draw[->, very thick] (l3)--(l4);
	\draw[->, very thick] (l4)--(l5);
	\draw[->, very thick] (l5)--(l6);
	\draw[->, very thick] (l6)--(l7);
	\draw[->, very thick] (l7)--(l8);
	\draw[->, very thick] (l8)--(l9);
	\draw[->, very thick] (l9)--(l10);
	\draw[->, very thick] (l10)--(l11);
	\draw[->, very thick] (l11)--(l12);
	\draw[->, very thick] (l12)--(l13);
	\draw[->, very thick] (l13)--(l14);
	\draw[->, very thick] (l14)--(output);
	
	\draw[->, very thick] (2.5cm,0) -- (2.5cm,2.5cm) -- (8cm,2.5cm) -- (8cm,2cm);
	\draw[->, very thick] (4.5cm,0) -- (4.5cm,2.25cm) -- (6cm,2.25cm) -- (6cm,2cm);
	
	\draw[->, very thick] (10.25cm,0) -- (10.25cm,2.5cm) -- (12cm,2.5cm) -- (12cm,2cm);
	\draw[->, very thick] (10.25cm,2.5cm) -- (14cm,2.5cm) -- (14cm,2cm);
\end{tikzpicture}
	\caption{Proposed neural speech codec architecture: Variables in brackets denote the output channels and number of output samples/frames.}
	\label{fig:NN_overview}
\end{figure*}
%
\section{\Revision{Proposed} Network Architecture}
\label{sec:network_architecture}
In the following section, we describe the proposed \ac{DNN} architecture for \Revision{real-time voice communication at low bitrates.} However, please note that the presented quantizers are applicable also to \Revision{other models and} general audio coding applications \Revision{(see also experiments in Sec.~\ref{sec:SQ_VQ_STFTvsFreqCodec})}. 

\Revision{The proposed neural speech codec operates in \ac{STFT} domain in order to emphasize harmonic structure of the generated signals and to simplify the training task by avoiding to learn an analysis frontend and a synthesis backend. To enable real-time operation, the proposed network architecture is causal and allows for realizations at very low computational complexity.}

\begin{figure*}
	\centering
	\subfloat[ConvBlock]{\begin{tikzpicture}
		\draw[dashed, very thick, gray] (0.5,-2.25)rectangle(5.5,2.25);
		
		\node[rotate=90](input) at (0,0){$\text{Input}(\NumCh, \NumFrames)$};
		\node[draw, thick, rotate=90, minimum width=4.25cm, below of = input](l1){$(\mathtt{BN} +)\ \mathtt{GELU}$};
		\node[draw, thick, rotate=90, minimum width=4.25cm, below of = l1](l2){$\text{Ch.-wise}\ \mathtt{Conv}_{\KernelSize}$};
		\node[draw, thick, rotate=90, minimum width=4.25cm, below of = l2](l3){$\mathtt{ChannelNorm}$};
		\node[draw, thick, rotate=90, minimum width=4.25cm, below of = l3](l4){$\ConvLin(\NumCh_{\text{up}}, \NumFrames) + \mathtt{GELU}$};
		\node[draw, thick, rotate=90, minimum width=4.25cm, below of = l4](l5){$\ConvLin(\NumCh, \NumFrames)$};
		\node[rotate=90, below of = l5](output){$\text{Output}(\NumCh, \NumFrames)$};
		
		\draw[->, very thick] (input)--(l1);
		\draw[->, very thick] (l1)--(l2);
		\draw[->, very thick] (l2)--(l3);
		\draw[->, very thick] (l3)--(l4);
		\draw[->, very thick] (l4)--(l5);
		\draw[->, very thick] (l5)--(output);
		
		\node at (0,-2.65){};
	\end{tikzpicture}
	\label{fig:NN_convblock}}
\hfill
\subfloat[EncBlock]{\begin{tikzpicture}
		\draw[dashed, very thick, gray, fill=blue!30!white] (0.5,-2.4)rectangle(3.5,2.4);
		
		\node[rotate=90](input) at (0,0){$\text{Input}(\NumCh_m, \NumFrames_m)$};
		\node[draw, thick, rotate=90, minimum width=4.6cm, below of = input](l1){$\mathtt{Conv}_{\KernelSize,\Stride}(\NumCh_{m+1},\NumFrames_{m+1})$};
		\node[draw, thick, rotate=90, minimum width=4.6cm, below of = l1](l2){$\text{ConvBlock}(2\NumCh_{m+1}, \NumFrames_{m+1})$};
		\node[draw, thick, rotate=90, minimum width=4.6cm, below of = l2](l3){$\text{ConvBlock}(\NumCh_{m+1}, \NumFrames_{m+1})$};
		\node[rotate=90, below of = l3](output){$\text{Output}(\NumCh_{m+1}, \NumFrames_{m+1})$};
		
		\draw[->, very thick] (input)--(l1);
		\draw[->, very thick] (l1)--(l2);
		\draw[->, very thick] (l2)--(l3);
		\draw[->, very thick] (l3)--(output);
		
		\node at (0,-2.5){};
	\end{tikzpicture}%
	\label{fig:NN_encblock}}
\hfill
\subfloat[DecBlock]{\begin{tikzpicture}
		\draw[dashed, very thick, gray, fill=yellow!30!white] (0.5,-2.2)rectangle(4.5,2.2);
		
		\node[rotate=90](input) at (0,0){$\text{Input}(\NumCh_{m}, \NumFrames_{m})$};
		\node[draw, thick, rotate=90, minimum width=4.2cm, below of = input](l1){$\text{ConvBlock}(2\NumCh_{m}, \NumFrames_{m})$};
		\node[draw, thick, rotate=90, minimum width=4.2cm, below of = l1](l2){$\text{ConvBlock}(\NumCh_{m}, \NumFrames_{m})$};
		\node[draw, thick, rotate=90, minimum width=4.2cm, below of = l2](l3){$\text{TADE}(\NumCh_{m}, \NumFrames_{m})$};
		\node[draw, thick, rotate=90, minimum width=4.2cm, below of = l3](l4){$\mathtt{ConvT}_{\KernelSize,\Stride}(\NumCh_{m+1}, \NumFrames_{m+1})$};
		\node[rotate=90, below of = l4](output){$\text{Output}(\NumCh_{m+1}, \NumFrames_{m+1})$};
		
		\node(cond) at (1,-2.5){$\text{Cond}(\NumCh_{\NumBlocks}, \NumFrames_{\NumBlocks})$};
		
		\draw[->, very thick] (input)--(l1);
		\draw[->, very thick] (l1)--(l2);
		\draw[->, very thick] (l2)--(l3);
		\draw[->, very thick] (l3)--(l4);
		\draw[->, very thick] (l4)--(output);
		
		\draw[->, very thick] (cond)-|(l3);
	\end{tikzpicture}%
	\label{fig:NN_decblock}}
	\caption{Convolutional network blocks: (a) Design of ConvBlocks ($\mathtt{BN}$ is not used for the EncBlocks) used in the (b) EncBlocks and (c) DecBlocks.}
	\label{fig:NN_convblocks}
\end{figure*}
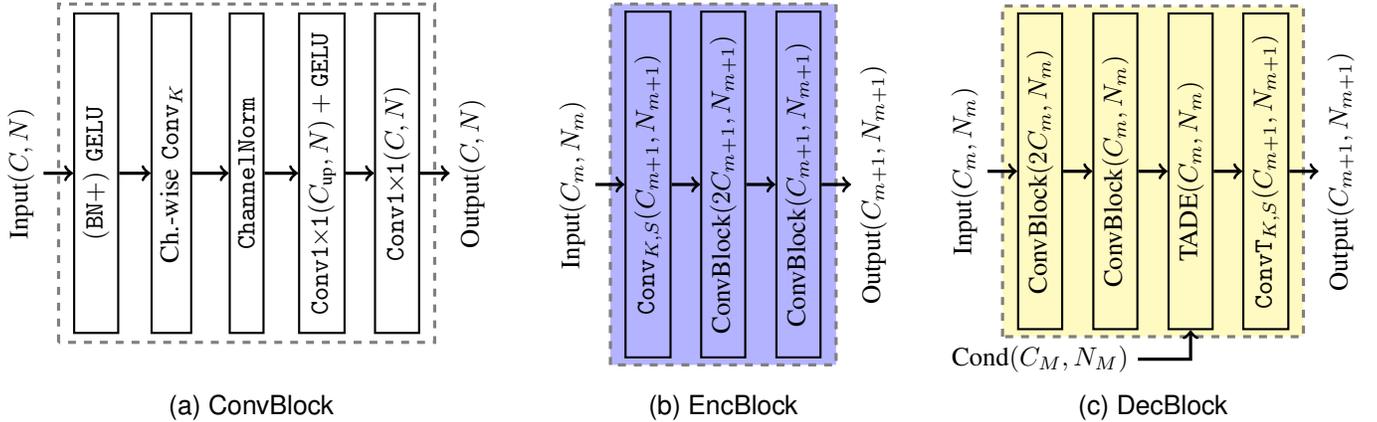

Its building blocks are illustrated in Figures \ref{fig:NN_overview}, \ref{fig:NN_convblocks}, \ref{fig:NN_dprnn}, \ref{fig:NN_tades}, where \ac{DNN} blocks are annotated with the corresponding output channels $\NumCh$ and number of (temporal) output samples or frames $\NumFrames$ if they are changed by the respective layer, otherwise we will omit this annotation for readability (we always omit the notation of the batch size $\BatchSize$ for the same reason). All convolutional layers in this paper are one-dimensional and causal and we denote them by $\mathtt{Conv}_{\KernelSize, \Stride, \Dilation}(\NumCh, \NumFrames)$, where $\KernelSize$, $\Stride$ and $\Dilation$ denote the kernel size, the stride and the dilation, respectively. To simplify notation, we omit $\Stride$ and $\Dilation$ if they are equal to one and denote convolution with $\KernelSize=1$ as $\ConvLin$. 
\subsection{Overview}
The \ac{DNN} architecture of the proposed neural speech codec is depicted in Fig.~\ref{fig:NN_overview}. The input signal is processed by an analysis block which is realized in this paper by the \ac{STFT}. The input signal waveform is transformed into a complex-valued \ac{STFT} representation $\STFTVar_{\FreqIdx,\FrameIdx}$, where $\FrameIdx$ and $\FreqIdx$ denote the frame and frequency index, respectively. To avoid dominance of high-energy components, the \ac{STFT} signals are compressed
\begin{equation}
	\vert \STFTVar_{\FreqIdx,\FrameIdx} \vert^{\alpha}\, e^{\jmath \angle \STFTVar_{\FreqIdx,\FrameIdx}},
\end{equation}
where $\alpha > 0$ is a compression factor (we use $0.3$ in the following), $\jmath$ is the imaginary unit and $\angle (\cdot)$ denotes the phase. Subsequently, the compressed \ac{STFT} signal is represented by real and imaginary part which are stacked along the frequency dimension, resulting in an input tensor $\mathbf{X}\in\mathbb{R}^{\BatchSize\times 2 F \times N}$ for the subsequent neural network block, where $\BatchSize$ denotes the batch size, $F$ the number of non-redundant \ac{STFT} bins and $N$ the number of frames. The feature representation outputted by the analysis block is successively downsampled by a cascade of encoder blocks (see Sec.~\ref{sec_nn:encoder_block}). \Revision{In the following, we treat the frequency axis as channel axis in the model, i.e., the encoder downsamples in time and in frequency direction.} Afterwards, a Dual-Path Convolutional Recurrent Neural Network (DPCRNN) layer (\Revision{may be considered as a variant of \cite{DPRNN},} see Sec.~\ref{sec_nn:dprnn}) is applied to exploit long-term signal relationships. To improve gradient flow and to effectively combine intermediate signal representations at the encoder before quantization (see Sec.~\ref{sec:vq} and Sec.~\ref{sec:sq}), \Revision{Temporal Adaptive DEnormalization (TADE) Residual Blocks (TADEResBlocks)} (see Sec.~\ref{sec_nn:tade}) are appended and conditioned on the outputs of the encoder blocks. The obtained quantized latent is processed by a decoder roughly symmetrical to the encoder, starting with a DPCRNN and followed by decoder blocks upsampling the latent successively. Here, each decoder block is conditioned on the quantized latent realizing a styling transform similar to StyleMelGAN \cite{mustafa_stylemelgan_2021}. The output of the final decoder block is processed by a synthesis block (\Revision{counterpart} to the analysis block) to reconstruct the time-domain signal waveform, where the inverse \ac{STFT} is preprocessd by the expansion
\begin{equation}
	\vert \STFTVarRec_{\FreqIdx,\FrameIdx} \vert^{\frac{1}{\alpha}}\, e^{\jmath \angle \STFTVarRec_{\FreqIdx,\FrameIdx}}.
\end{equation}
\subsection{Encoder Block}
\label{sec_nn:encoder_block}
\Revision{As shown in Fig.~\ref{fig:NN_convblocks} (b), each} encoder block potentially downsamples the input signal \Revision{in time direction} by a strided convolutional layer that may also adjust the number of channels\Revision{, i.e., reduce the frequency resolution}. The result is further processed by two ConvBlocks where the first one doubles the number of channels and the second restores the number of input channels again. Both ConvBlocks do not alter the time resolution.

\Revision{As can be seen in Fig.~\ref{fig:NN_convblocks} (a), each} ConvBlock applies a $\mathtt{GELU}$ activation to the input (for the encoder blocks, no Batch Normalization $\mathtt{BN}$ is used), which is followed by a channel-wise convolution with kernel size $\KernelSize$ and same number of output channels $\NumCh$\Revision{, i.e., maintaining the same frequency resolution}. \Revision{Here, individual filters are learned for each channel.} Then, $\mathtt{ChannelNorm}$ is followed by a $\ConvLin$ increasing the channels to $\NumCh_{\text{up}}$ and a $\mathtt{GELU}$ activation. Finally, the channels are again changed to the number of input channels by another $\ConvLin$. \Revision{These two $\ConvLin$ layers couple the output of the channel-wise filtering.}
\subsection{DPCRNN}
\label{sec_nn:dprnn}
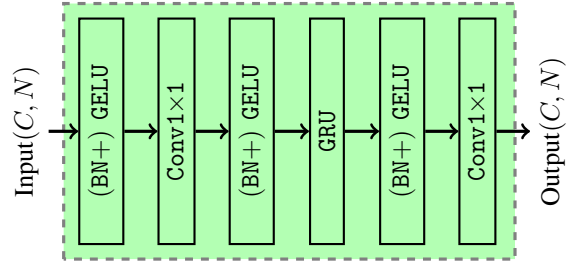
\begin{figure}
	\centering
	\begin{tikzpicture}		
	\draw[dashed, very thick, gray, fill=green!30!white] (0.5,-1.7)rectangle(6.5,1.7);
	
	\node[rotate=90](input) at (0,0){$\text{Input}(\NumCh, \NumFrames)$};
	\node[draw, thick, rotate=90, minimum width=3cm, below of = input](l1){$(\mathtt{BN} +)\ \mathtt{GELU}$};
	\node[draw, thick, rotate=90, minimum width=3cm, below of = l1](l2){$\ConvLin$};
	\node[draw, thick, rotate=90, minimum width=3cm, below of = l2](l3){$(\mathtt{BN} +)\ \mathtt{GELU}$};
	\node[draw, thick, rotate=90, minimum width=3cm, below of = l3](l4){$\mathtt{GRU}$};
	\node[draw, thick, rotate=90, minimum width=3cm, below of = l4](l5){$(\mathtt{BN} +)\ \mathtt{GELU}$};
	\node[draw, thick, rotate=90, minimum width=3cm, below of = l5](l6){$\ConvLin$};
	\node[rotate=90, below of = l6](output){$\text{Output}(\NumCh, \NumFrames)$};
	
	\draw[->, very thick] (input)--(l1);
	\draw[->, very thick] (l1)--(l2);
	\draw[->, very thick] (l2)--(l3);
	\draw[->, very thick] (l3)--(l4);
	\draw[->, very thick] (l4)--(l5);
	\draw[->, very thick] (l5)--(l6);
	\draw[->, very thick] (l6)--(output);
\end{tikzpicture}
	\caption{DPCRNN network block ($\mathtt{BN}$ is not used on encoder side).}
	\label{fig:NN_dprnn}
\end{figure}
\Revision{As shown in Fig.~\ref{fig:NN_dprnn}, the} DPCRNN normalizes the input by $\mathtt{BN}$ (only on decoder side, on encoder side all $\mathtt{BN}$ are dropped) and applies a $\mathtt{GELU}$ activation. Channels are shuffled by $\ConvLin$ and the result is normalized by $\mathtt{BN}$ and activated by $\mathtt{GELU}$. A $\mathtt{GRU}$ layer is used for modeling temporal context, which is followed by another $\mathtt{BN}$, $\mathtt{GELU}$ and $\ConvLin$.
\subsection{TADE and TADEResBlock}
\label{sec_nn:tade}
\begin{figure*}
	\centering
	\subfloat[TADE]{\begin{tikzpicture}
		\draw[dashed, very thick, gray] (-5.5,-3.5)rectangle(1.5,-0.5);
		
		\node(input) at (0,0){$\text{Input}(\NumCh, \NumFrames)$};
		\node[draw, thick, minimum width=2.75cm, below of = input](r1){$\mathtt{ChannelNorm}$};
		\node[below of = r1](r2){{\Large$\bm{\odot}$}};
		\node[below of = r2](r3){{\Large$\bm{\oplus}$}};
		
		\node[below of = r3](output){$\text{Output}(\NumCh, \NumFrames)$};
		
		\node(cond) at (-4,0){$\text{Cond}(\tilde{\NumCh}, \tilde{\NumFrames})$};
		\node[draw, thick, minimum width=2.75cm, below of = cond](l1){$\mathtt{Up}/\mathtt{Down}(\tilde{\NumCh}, \NumFrames)$};
		\node[draw, thick, minimum width=2.75cm, below of = l1](l2){$\mathtt{Conv}_{\KernelSize}(\NumCh, \NumFrames)$};
		\node[draw, thick, minimum width=2.75cm, below of = l2](l3){$\mathtt{LeakyReLU(0.2)}$};
		\node[draw, thick, right of = l3, yshift=1cm, xshift=1.5cm](l4a){$\mathtt{Conv}_{\KernelSize}$};
		\node[draw, thick, right of = l3, yshift=0cm, xshift=1.5cm](l4b){$\mathtt{Conv}_{\KernelSize}$};
		
		\draw[->, very thick] (cond)--(l1);
		\draw[->, very thick] (l1)--(l2);
		\draw[->, very thick] (l2)--(l3);
		\draw[->, very thick] (-2.4,-3)|-(l4a);
		\draw[->, very thick] (l3)--(l4b);
		\draw[->, very thick] (l4a)--(r2);
		\draw[->, very thick] (l4b)--(r3);
		
		\draw[->, very thick] (input)--(r1);
		\draw[->, very thick] (r1)--(r2);
		\draw[->, very thick] (r2)--(r3);
		\draw[->, very thick] (r3)--(output);
	\end{tikzpicture}
	\label{fig:NN_tade}}
\hfill
\subfloat[TADEResBlock]{\begin{tikzpicture}
		\draw[dashed, very thick, gray, fill=red!30!white] (0.25,-2.05)rectangle(9.05,2.05);
		
		\node[rotate=90](input) at (-0.2,0){$\text{Input}(\NumCh, \NumFrames)$};
		\node[draw, thick, rotate=90, minimum width=3cm, below of = input, yshift=-0.25cm](r1){TADE};
		\node[draw, thick, rotate=90, below left of = r1, yshift=-0.25cm, xshift=-0.2cm](r2a){$\mathtt{Conv}_{\KernelSize}$};
		\node[draw, thick, rotate=90, below right of = r1, yshift=-0.25cm, xshift=0.2cm](r2b){$\mathtt{Conv}_{\KernelSize}$};
		\node[draw, thick, rotate=90, below of = r2a](r3a){$\mathtt{tanh}$};
		\node[draw, thick, rotate=90, below of = r2b](r3b){$\mathtt{softmax}$};
		\node[below of = r1, rotate=90, yshift=-2.75cm, xshift=1cm](r4){{\Large$\bm{\odot}$}};
		
		\node[draw, thick, rotate=90, minimum width=3cm, below of = r4](r5){TADE};
		\node[draw, thick, rotate=90, below left of = r5, yshift=-0.25cm, xshift=-0.2cm](r6a){$\mathtt{Conv}_{\KernelSize,\Dilation}$};
		\node[draw, thick, rotate=90, below right of = r5, yshift=-0.25cm, xshift=0.2cm](r6b){$\mathtt{Conv}_{\KernelSize,\Dilation}$};
		\node[draw, thick, rotate=90, below of = r6a](r7a){$\mathtt{tanh}$};
		\node[draw, thick, rotate=90, below of = r6b](r7b){$\mathtt{softmax}$};
		\node[below of = r5, rotate=90, yshift=-2.75cm, xshift=1cm](r8){{\Large$\bm{\odot}$}};
		\node[right of = r8](r9){{\Large$\bm{\oplus}$}};		
		\node[right of = r9, rotate=90](output){$\text{Output}(\NumCh, \NumFrames)$};
		\node(cond) at (1.025,-2.6){$\text{Cond}(\tilde{\NumCh}, \tilde{\NumFrames})$};
		
		\draw[->, very thick] (cond)--(r1);
		\draw[->, very thick] (1.025,-1.8)-|(r5);
		\draw[->, very thick] (input)--(r1);
		\draw[->, very thick] (0.5,0)--(0.5,1.8)-|(r9);
		\draw[->, very thick] (r1)--(r2a);
		\draw[->, very thick] (r1)--(r2b);
		\draw[->, very thick] (r2a)--(r3a);
		\draw[->, very thick] (r2b)--(r3b);
		\draw[->, very thick] (r3a)--(r4);
		\draw[->, very thick] (r3b)--(r4);
		
		\draw[->, very thick] (r4)--(r5);
		\draw[->, very thick] (r5)--(r6a);
		\draw[->, very thick] (r5)--(r6b);
		\draw[->, very thick] (r6a)--(r7a);
		\draw[->, very thick] (r6b)--(r7b);
		\draw[->, very thick] (r7a)--(r8);
		\draw[->, very thick] (r7b)--(r8);
		\draw[->, very thick] (r8)--(r9);
		\draw[->, very thick] (r9)--(output);
	\end{tikzpicture}%
	\label{fig:NN_taderesblock}}
	\caption{(a) The TADE layer is a component of the DecBlocks as well fo the (b) TADEResBlock, which consists of two TADE layers with gated activations and a residual connection.}
	\label{fig:NN_tades}
\end{figure*}
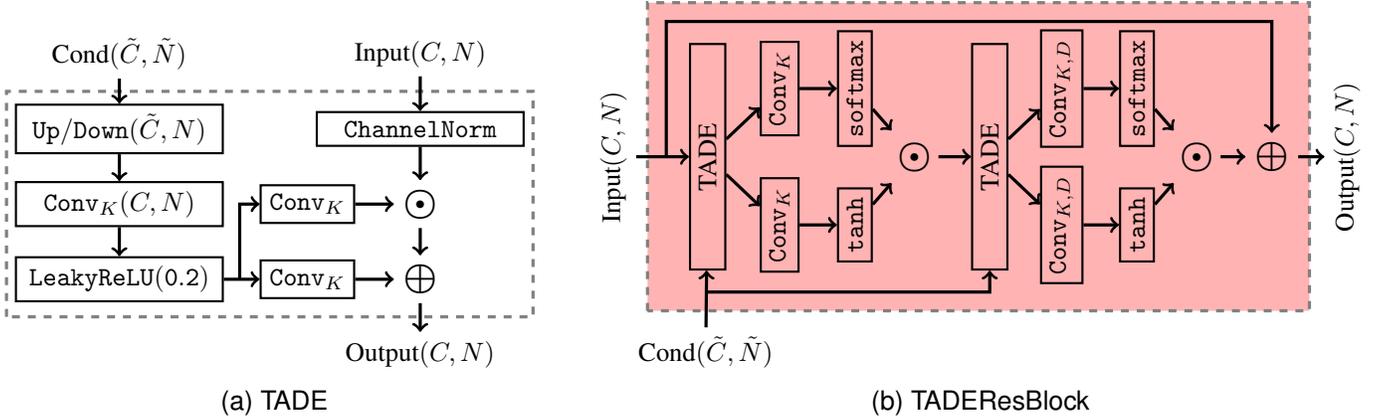
The encoder output is styled by a sequence of TADEResBlocks \Revision{(see Fig.~\ref{fig:NN_tades} (b))} that are conditioned on the outputs of the EncBlocks. The input of the TADEResBlock is styled by a TADE layer \Revision{(see Fig.~\ref{fig:NN_tades} (a))} \cite{mustafa_stylemelgan_2021} and activated by $\mathtt{softmax}$-gated $\mathtt{tanh}$ activation, i.e., is processed in parallel by $\mathtt{conv}$ layers one with $\mathtt{softmax}$ and the other one with $\mathtt{tanh}$ activation where the results are multiplied together. This structure is repeated where the $\mathtt{conv}$ layers of the gated activation are now parameterized with dilation. Both structures together are then realized as a ResBlock.

The \Revision{TADE} layer is a building block of the TADEResBlocks \Revision{(see Fig.~\ref{fig:NN_tades} (b))} as well as for the DecBlocks \Revision{(see Fig.~\ref{fig:NN_convblocks} (c))}. It normalizes the input by ChannelNorm and adjusts it by an affine transform. The scale and shift parameters of the affine transform are learned from a conditioning signal that is preprocessed by up- or downsampling and a $\mathtt{conv}$ layer activated by $\mathtt{LeakyReLU}$. From this, feature-wise scale and shift parameters are calculated by $\mathtt{conv}$ layers.
\subsection{Decoder Block}
\label{sec_nn:decoder_block}
Similar as in StyleMelGAN \cite{mustafa_stylemelgan_2021} the latent is styled by a sequence of decoder network blocks conditioned on the quantized latent. Similar to the EncBlocks, the signal is processed by two ConvBlocks, where the first one doubles the input channels and the second restores the number of input channels again. Differently to its use in the EncBlocks, the ConvBlocks in the DecBlocks use $\mathtt{BN}$. The output of the ConvBlocks\Revision{, which perform a channel-wise filtering and then a combination of the result (cf. Sec.~\ref{sec_nn:encoder_block}),} is styled by a TADE layer conditioned on the quantized latent. Finally, a transposed conv layer $\mathtt{ConvT}$ upsamples the signal.
\subsection{\Revision{Relation to Prior Art}}
\label{sec_nn:rel_prior_art}
\Revision{Many of previously proposed neural audio codecs are based on the SEANet architecture \cite{tagliasacchi_seanet_2020}, e.g., Soundstream \cite{zeghidour_soundstream_2021}, EnCodec \cite{defossez_high_2022} or AudioDec \cite{wu_audiodec_2023}. Here, the input signal is downsampled in the encoder in order to obtain a compact representation in the bottleneck for quantization and is upsampled on decoder side in a mirrored fashion. In this case, more and more feature channels are extracted from the input signals at each downsampling step. Our proposed architecture has two main novel distinctive features. First, the use of conditioning layers (TADEResBlocks and TADE layers) for an intelligent implementation of skip connections at encoder and decoder side. Second, the overall processing scheme of the input data: Instead of extracting more and more feature channels at each downsampling step, we treat the frequency axis as channels resulting in a pure upsampling/downsampling architecture. In this way, also a structural bias is imposed on the network, i.e., the outer blocks of the network dominate complexity and model the fine structure of the signals, whereas the network blocks that are closer to the quantizer model are much less computationally complex and model more high-level features of the waveform. The separate processing of temporal and frequency-specific features allows for more efficient realizations than other state-of-the-art approaches (see Sec.~\ref{sec:experiments}).}
\subsection{Losses}
\label{sec:losses}
For training the neural speech codec, we use several loss functions including a reconstruction loss operating in the \ac{STFT} domain. Here, we denote the \ac{STFT} representation of the target signal by $\STFT_{i}\in\mathbb{C}^{\NumFreqs_{i} \times \NumFrames_i}$ and the \ac{STFT} of the generated signal by $\STFTRec_{i}\in\mathbb{C}^{\NumFreqs_i \times \NumFrames_i}$. We compare multiple \ac{STFT} representations with the reconstruction loss at different resolutions which are denoted by $i\in \mathcal{R}$. With these definitions, the reconstruction loss is denoted as \cite{yamamoto_parallel_2020}
\begin{equation}
	\mathcal{L}_\text{rec} = \frac{1}{\vert \mathcal{R}\vert} \sum_{i\in\mathcal{R}}\Expect_{p(\XVec)}\left[\mathcal{L}_\text{MSE}\left(\STFT_{i}, \STFTRec_{i}\right) + \mathcal{L}_\text{mag}\left(\STFT_{i}, \STFTRec_{i}\right)\right],
\end{equation}
with the relative \ac{MSE} loss component
\begin{equation}
	\mathcal{L}_\text{MSE}\left(\STFT_{i}, \STFTRec_{i}\right) = \frac{\left\Vert \STFT_{i} - \STFTRec_{i}\right\Vert_{\text{F}}}{\left\Vert \STFT_{i} \right\Vert_{\text{F}}}
\end{equation}
and the logarithmic absolute error component
\begin{equation}
	\mathcal{L}_\text{mag}\left(\STFT_{i}, \STFTRec_{i}\right) =  \left\Vert \log \STFT_{i} - \log \STFTRec_{i}\right\Vert_{1,1}.
\end{equation}
Here, $\Vert \cdot\Vert_{1,1}$ denotes the sum of absolute values of matrix elements, i.e., corresponding to an $\ell 1$ \Revision{loss.}

We use an ensemble of three identical but differently initialized discriminators $\mathtt{D}_k$, $k\in\{1,2,3\}$ based on the discriminator architecture proposed \Revision{in MelGAN} \cite{kumar_melgan_2019}. The adversarial loss component for training the generator $\mathtt{G}$, i.e., the neural speech codec, \Revision{is} based on the \ac{LS} \ac{GAN} loss \cite{mao_least_2017}
\begin{equation}
	\mathcal{L}_\text{adv} = \Expect_{p(\XVec)}\left[\sum_{k=1}^{3} \left(\mathtt{D}_k\left(\mathtt{G}(\XVec)\right)-1\right)^2\right].
\end{equation}
By learning to distinguish between unprocessed and coded items, the discriminator extracts meaningful intermediate feature representations $\mathtt{D}^{(j)}_k(\XVec)$, that are compared in an additional feature matching loss
\begin{equation}
	\mathcal{L}_\text{feat} = \frac{1}{3J}\ \Expect_{p(\XVec)}\left[\sum_{k=1}^{3}\sum_{j=1}^{J} \left\Vert \mathtt{D}^{(j)}_k(\XVec) - \mathtt{D}^{(j)}_k\left(\mathtt{G}(\XVec)\right)\right\Vert_{1,1}\right].
\end{equation}
Here, $j\in\{1,\dots,J\}$ indexes the intermediate feature representation.

Finally, all discussed loss functions are combined to the training loss $\mathcal{L}$ by a weighted sum
\begin{equation}
	\mathcal{L}_{\text{Gen}} = \lambda_\text{rec}\mathcal{L}_\text{rec} + \lambda_\text{adv}\mathcal{L}_\text{adv} + \lambda_\text{feat}\mathcal{L}_\text{feat}.
	\label{eq:loss_weighted_sum}
\end{equation}
Here, $\lambda_\text{rec}, \lambda_\text{adv}$ and $\lambda_\text{feat}$ are nonnegative weighting factors. Note that for \ac{VQ} we may have an additional commitment loss term in \eqref{eq:loss_weighted_sum}.

The discriminator is also trained with an \ac{LS} \ac{GAN} loss, corresponding to
\begin{equation}
	\mathcal{L}_{\text{Disc}} = \Expect_{p(\XVec)}\left[\sum_{k=1}^{3} \big(\mathtt{D}_k\left(\XVec\right) - 1\big)^2 + \big(\mathtt{D}_k\left(\mathtt{G}(\XVec)\right)\big)^2\right].
\end{equation}
\section{Experiments}
\label{sec:experiments}
\begin{table}
	\centering
	\begin{tabular}{rcccccc}
		\toprule
		Block index $m$ & $1$ & $2$ & $3$ & $4$ & $5$ & $6$ \\
		\midrule
		Kernel size $\KernelSize$& $7$ & $5$ & $5$ & $5$ & $3$ & $3$ \\
		Num. channels $C$& $256$ & $128$ & $64$ & $64$ & $32$ & $32$ \\
		Stride $\Stride$& $1$ & $1$ & $1$ & $1$ & $1$ & $2$ \\
		\bottomrule
	\end{tabular}\vspace{3pt}
	\caption{Parameters used for experiments with the proposed neural speech codec \Revision{($\NumBlocks=6$ encoder/decoder blocks)}.}
	\label{tab:params_dnn}
\end{table}
In the following, we present experimental results for the proposed quantization method and the proposed neural speech codec.
\subsection{Experimental Setup}
The proposed neural speech codec was trained with the full VCTK dataset comprising 44 hours \cite{veaux_superseded_2017} and 260 hours of clean speech signals from the LibriTTS dataset \cite{zen_libritts_2019}. All processing is done at a sampling rate of $16\,\mathrm{kHz}$. Both neural codec and \Revision{discriminators} are trained for $700\,\mathrm{k}$ iterations on $2\,s$ long signal chunks with the AdamW optimizer \cite{loshchilov_decoupled_2019}, a learning rate of $0.001$, a batch size of $\BatchSize=128$ and an annealed cosine learning rate scheduler with a linear warm up phase. The training is performed in two phases: In a first phase, the codec is trained without discriminator ($\lambda_\text{adv}=\lambda_\text{feat}=0$) and $\lambda_\text{rec}=1$ (see \eqref{eq:loss_weighted_sum}). After $100$ epochs, the \Revision{discriminators} enters training by setting $\lambda_\text{adv}=1$ $\lambda_\text{feat}=10$. For realizing the analysis transform, we use for the \ac{STFT} a Hann window of $320$ samples length and $160$ samples \Revision{stride}. The resulting signal blocks are zero-padded to an FFT length of $512$ samples. The proposed neural speech codec is parameterized according to Tab.~\ref{tab:params_dnn} in the following experiments.

We use a test set compiled from various sources (including the \Revision{ITU-T} Standard P.501 (20/05), ETSI TS 103 281 Annex E and the NTT dataset), which comprises 28 items from English speakers (evenly split between female and male) which is chosen to be diverse in speaking style, content and speaker identity. For objective evaluation, we use \Revision{VISQOL} \cite{chinen_visqol_2020} and ESTOI \cite{jensen_algorithm_2016} were larger values correspond to better performance as objective measures. Furthermore, we employ NOMAD \cite{ragano_nomad_2024} were we only evaluate matching generated and ground truth signals (full-reference mode). Here, lower values correspond to superior performance. In our assessment, NOMAD reflected the perceived audio quality best.

To asses perceived signal quality, we performed crowd-sourced listening test with naive listeners according to the P.808 standard \cite{noauthor_recommendation_2021} with AWS Mechanical Turk. All conditions of the listening tests produce wide-band speech at $16\,\mathrm{kHz}$. The listeners are post screened according to trap questions and their scoring of the Modulated Noise Reference Unit (MNRU) conditions, i.e., anchor signals artificially generated with controlled degradations. 
%
\subsection{\Revision{Effect of Projection}}
\begin{figure}
\begin{tikzpicture}

\definecolor{darkgray176}{RGB}{176,176,176}
\definecolor{darkorange25512714}{RGB}{255,127,14}
\definecolor{forestgreen4416044}{RGB}{44,160,44}

\begin{axis}[
width=0.27\textwidth,
height=0.3\textheight,
y grid style={gray},
ymajorgrids,
tick align=outside,
tick pos=left,
x grid style={darkgray176},
xmin=0.5, xmax=2.5,
xtick style={color=black},
y grid style={darkgray176},
ymin=0.13, ymax=0.95,
ytick style={color=black},
xtick={1,2},
xticklabels = {$\bar{\rho}_{\mathbf{Y}, \text{tmp}}$, $\bar{\rho}_{\mathbf{Z}, \text{tmp}}$},
]
\addplot [black, very thick]
table {%
0.9 0.229443430500113
1.1 0.229443430500113
1.1 0.266928127866114
0.9 0.266928127866114
0.9 0.229443430500113
};
\addplot [black, very thick]
table {%
1 0.229443430500113
1 0.206226937385638
};
\addplot [black, very thick]
table {%
1 0.266928127866114
1 0.319891650103174
};
\addplot [black, very thick]
table {%
0.9625 0.206226937385638
1.0375 0.206226937385638
};
\addplot [black, very thick]
table {%
0.9625 0.319891650103174
1.0375 0.319891650103174
};
\addplot [black, very thick, mark=o, mark size=3, mark options={solid,fill opacity=0}, only marks]
table {%
1 0.324461869699152
};
\addplot [black, very thick]
table {%
1.9 0.168566549836326
2.1 0.168566549836326
2.1 0.220649818359225
1.9 0.220649818359225
1.9 0.168566549836326
};
\addplot [black, very thick]
table {%
2 0.168566549836326
2 0.138700686217842
};
\addplot [black, very thick]
table {%
2 0.220649818359225
2 0.277272280491631
};
\addplot [black, very thick]
table {%
1.9625 0.138700686217842
2.0375 0.138700686217842
};
\addplot [black, very thick]
table {%
1.9625 0.277272280491631
2.0375 0.277272280491631
};
\addplot [black, very thick, mark=o, mark size=3, mark options={solid,fill opacity=0}, only marks]
table {%
2 0.305953423674042
2 0.305574170908764
2 0.3111992856503
};
\addplot [black, very thick]
table {%
0.9 0.250433417324596
1.1 0.250433417324596
};
\addplot [black, very thick, mark=triangle*, mark size=3, mark options={solid}, only marks]
table {%
1 0.252130263488542
};
\addplot [black, very thick]
table {%
1.9 0.194721760810209
2.1 0.194721760810209
};
\addplot [black, very thick, mark=triangle*, mark size=3, mark options={solid}, only marks]
table {%
2 0.200900101279952
};
\end{axis}

\end{tikzpicture}
\begin{tikzpicture}

\definecolor{darkgray176}{RGB}{176,176,176}
\definecolor{darkorange25512714}{RGB}{255,127,14}
\definecolor{forestgreen4416044}{RGB}{44,160,44}

\begin{axis}[
width=0.27\textwidth,
height=0.3\textheight,
y grid style={gray},
ymajorgrids,
tick align=outside,
tick pos=left,
x grid style={darkgray176},
xmin=0.5, xmax=2.5,
xtick style={color=black},
y grid style={darkgray176},
ymin=0.13, ymax=0.95,
ytick style={color=black}, 
xtick={1,2},
xticklabels = {$\bar{\rho}_{\mathbf{Y}, \text{ch}}$, $\bar{\rho}_{\mathbf{Z}, \text{ch}}$},
]
\addplot [black, very thick]
table {%
0.9 0.873271093827523
1.1 0.873271093827523
1.1 0.903615563309355
0.9 0.903615563309355
0.9 0.873271093827523
};
\addplot [black, very thick]
table {%
1 0.873271093827523
1 0.85860977503665
};
\addplot [black, very thick]
table {%
1 0.903615563309355
1 0.922313678027357
};
\addplot [black, very thick]
table {%
0.9625 0.85860977503665
1.0375 0.85860977503665
};
\addplot [black, very thick]
table {%
0.9625 0.922313678027357
1.0375 0.922313678027357
};
\addplot [black, very thick]
table {%
1.9 0.301263929733083
2.1 0.301263929733083
2.1 0.369496666787245
1.9 0.369496666787245
1.9 0.301263929733083
};
\addplot [black, very thick]
table {%
2 0.301263929733083
2 0.293756362335816
};
\addplot [black, very thick]
table {%
2 0.369496666787245
2 0.470111140165606
};
\addplot [black, very thick]
table {%
1.9625 0.293756362335816
2.0375 0.293756362335816
};
\addplot [black, very thick]
table {%
1.9625 0.470111140165606
2.0375 0.470111140165606
};
\addplot [black, very thick]
table {%
0.9 0.883546363696507
1.1 0.883546363696507
};
\addplot [black, mark=triangle*, mark size=3, mark options={solid}, only marks]
table {%
1 0.886270520300041
};
\addplot [black, very thick]
table {%
1.9 0.326855576873136
2.1 0.326855576873136
};
\addplot [black, mark=triangle*, mark size=3, mark options={solid}, only marks]
table {%
2 0.347553049614055
};
\end{axis}

\end{tikzpicture}
	\caption{\Revision{Average correlation coefficients for neural encoder output before $\mathbf{Y}$ and after $\mathbf{Z}$ projection across time (left) and latent channels (right). The black triangles mark the mean of the results.}}
	\label{fig:correlations_projection}
\end{figure}
\Revision{In this section, we investigate the effect of the projection applied to the output of the encoder $\LatVec_\FrameIdx = \Encoder(\XVec_\FrameIdx) \in \mathbb{R}^\DimLatent$ which yields $\SQVec_\FrameIdx\in\mathbb{R}^\NumQuants$. To this end, we evaluate the average correlation coefficients across time and across latent dimensions before and after projection. In this paper, the projector consists of a one-dimensional convolution with kernel size of $3$ and a subsequent linear layer performing dimensionality reduction from $\DimLatent$ to $\NumQuants$. The output frames of the neural encoder and the resulting frames after projection $\SQVec_\FrameIdx=\mathtt{proj}(\LatVec_\FrameIdx)$ are stacked in the data matrices
\begin{equation}
	\mathbf{Y} = \left[\LatVec_1, \dots, \LatVec_\NumFrames\right] \in \mathbb{R}^{\DimLatent\times\NumFrames}
\end{equation}
and
\begin{equation}
	\mathbf{Z} = \left[\SQVec_1,\dots,\SQVec_\NumFrames\right] \in \mathbb{R}^{\NumQuants\times\NumFrames},
\end{equation}
respectively. For calculating the temporal correlation of the encoder output, we center the corresponding data matrix (i.e., we subtract the channel-wise mean $\frac{1}{\NumFrames}\mathbf{Y}\mathbf{1}$, where $\mathbf{1}\in\mathbb{R}^{\NumFrames}$ denotes the all-ones vector)
\begin{equation}
	\tilde{\mathbf{Y}}_{\text{tmp}} = \mathbf{Y} - \frac{1}{\NumFrames}\mathbf{Y}\mathbf{1}\mathbf{1}\transp
	\label{eq:mean_removal}
\end{equation}
and calculate its outer product
\begin{equation}
	\bm{\Sigma}_{\mathbf{Y},\text{tmp}}= \frac{1}{\NumFrames-1}\tilde{\mathbf{Y}}_{\text{tmp}}\left(\tilde{\mathbf{Y}}_{\text{tmp}}\right)\transp.
	\label{eq:covmat_estimate}
\end{equation}
To obtain the average correlation, we calculate the correlation coefficient element-wise and average over the magnitudes of the off-diagonal elements of the resulting matrix
\begin{equation}
	\bar{\rho}_{\mathbf{Y},\text{tmp}} = \sum_{\substack{i,j=1\\ i\neq j}}^{\DimLatent} \left\vert \frac{\bm{\Sigma}_{\mathbf{Y},ij,\text{tmp}}}{\sqrt{\bm{\Sigma}_{\mathbf{Y},ii,\text{tmp}}\bm{\Sigma}_{\mathbf{Y},jj,\text{tmp}}}}\right\vert \in [0,1].
	\label{eq:average_corrcoef}
\end{equation}
The calculation of the average correlation across channels $\bar{\rho}_{\mathbf{Y}, \text{ch}}$ is computed analogously by using transposed matrices in \eqref{eq:mean_removal}, adjusting the prefactor in \eqref{eq:covmat_estimate} and the summation in \eqref{eq:average_corrcoef}. The corresponding statistics after projection $\bar{\rho}_{\mathbf{Z}, \text{tmp}}$ and $\bar{\rho}_{\mathbf{Z},\text{ch}}$ are calculated the same way by replacing the outputs of the neural encoder $\mathbf{Y}$ with its projections $\mathbf{Z}$.
}

\Revision{The results can be seen in Fig.~\ref{fig:correlations_projection}, where the average correlation coefficient has been evaluated over the test set for the proposed neural codec architecture trained with \ac{SQ} with noise addition at $1.5\,\mathrm{kbps}$. The temporal correlation is slightly reduced by the projection, which is attributed to the use of a one-dimensional convolution in the projector. Much more pronounced is the reduction in correlation across the dimensions of the latent representation, i.e., the projector significantly decorrelates the latent frames before presented to the quantizer. This is consistent with classical coding principles where signal transforms are employed to decorrelate the input before scalar quantization.
}
%
\subsection{Ablation Study}
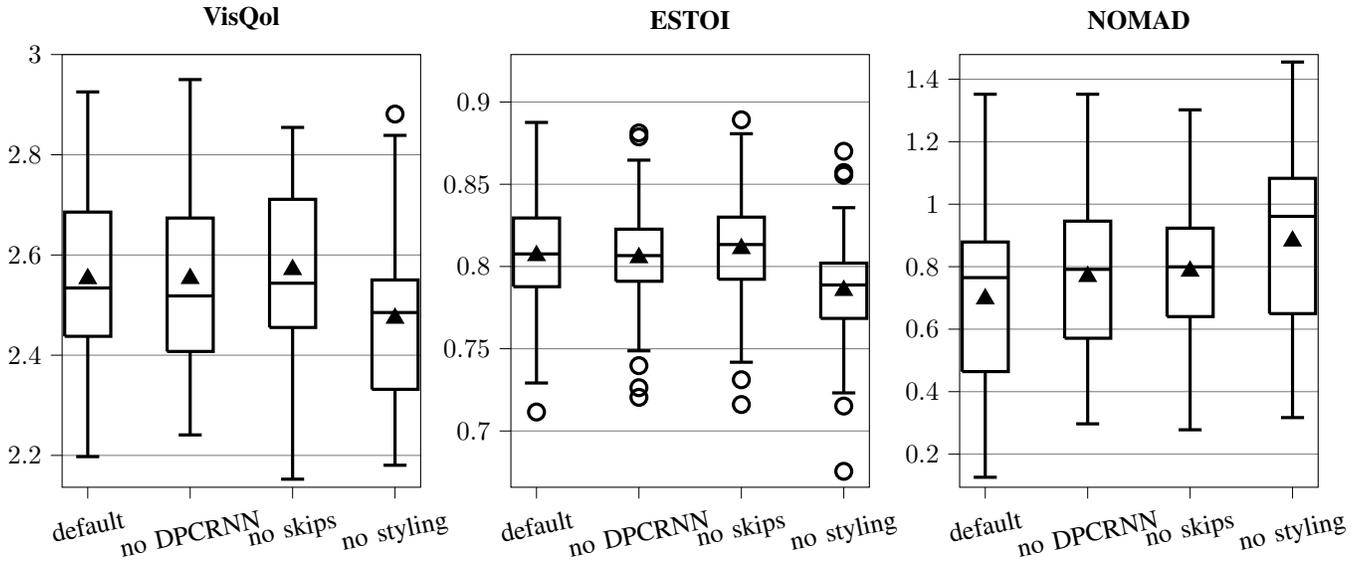
\begin{figure*}
\begin{tikzpicture}

\begin{groupplot}[group style={group size=3 by 1, horizontal sep=1.2cm}]
\nextgroupplot[
tick align=outside,
tick pos=left,
width = 0.35\textwidth,
height = 0.3\textheight,
title={\textbf{VisQol}},
x grid style={gray},
xmin=0.75, xmax=4.25,
xtick style={color=black},
xtick={1,2,3,4},
xticklabels={default, no DPCRNN, no skips, no styling},
xticklabel style = {rotate=12},
y grid style={gray},
ymajorgrids,
ymin=2.136452411, ymax=3,
ytick style={color=black}
]
\addplot [black, very thick]
table {%
	0.775 2.4376703475
	1.225 2.4376703475
	1.225 2.6855823525
	0.775 2.6855823525
	0.775 2.4376703475
};
\addplot [black, very thick]
table {%
	1 2.4376703475
	1 2.19755392
};
\addplot [black, very thick]
table {%
	1 2.6855823525
	1 2.92543284
};
\addplot [black, very thick]
table {%
	0.8875 2.19755392
	1.1125 2.19755392
};
\addplot [black, very thick]
table {%
	0.8875 2.92543284
	1.1125 2.92543284
};
\addplot [black, very thick]
table {%
	1.775 2.4074579275
	2.225 2.4074579275
	2.225 2.6737165875
	1.775 2.6737165875
	1.775 2.4074579275
};
\addplot [black, very thick]
table {%
	2 2.4074579275
	2 2.24091356
};
\addplot [black, very thick]
table {%
	2 2.6737165875
	2 2.95006553
};
\addplot [black, very thick]
table {%
	1.8875 2.24091356
	2.1125 2.24091356
};
\addplot [black, very thick]
table {%
	1.8875 2.95006553
	2.1125 2.95006553
};
\addplot [black, very thick]
table {%
	2.775 2.4553269225
	3.225 2.4553269225
	3.225 2.7110547525
	2.775 2.7110547525
	2.775 2.4553269225
};
\addplot [black, very thick]
table {%
	3 2.4553269225
	3 2.15254244
};
\addplot [black, very thick]
table {%
	3 2.7110547525
	3 2.85457739
};
\addplot [black, very thick]
table {%
	2.8875 2.15254244
	3.1125 2.15254244
};
\addplot [black, very thick]
table {%
	2.8875 2.85457739
	3.1125 2.85457739
};
\addplot [black, very thick]
table {%
	3.775 2.3318674625
	4.225 2.3318674625
	4.225 2.550222425
	3.775 2.550222425
	3.775 2.3318674625
};
\addplot [black, very thick]
table {%
	4 2.3318674625
	4 2.18036063
};
\addplot [black, very thick]
table {%
	4 2.550222425
	4 2.83871229
};
\addplot [black, very thick]
table {%
	3.8875 2.18036063
	4.1125 2.18036063
};
\addplot [black, very thick]
table {%
	3.8875 2.83871229
	4.1125 2.83871229
};
\addplot [black, very thick, mark=o, mark size=3, mark options={solid,fill opacity=0}, only marks]
table {%
	4 2.88123651
};
\addplot [black, very thick]
table {%
	0.775 2.534211235
	1.225 2.534211235
};
\addplot [black, very thick, mark=triangle*, mark size=3, mark options={solid}, only marks]
table {%
	1 2.55276680325
};
\addplot [black, very thick]
table {%
	1.775 2.51833248
	2.225 2.51833248
};
\addplot [black, very thick, mark=triangle*, mark size=3, mark options={solid}, only marks]
table {%
	2 2.55306369075
};
\addplot [black, very thick]
table {%
	2.775 2.543656075
	3.225 2.543656075
};
\addplot [black, very thick, mark=triangle*, mark size=3, mark options={solid}, only marks]
table {%
	3 2.5709188465
};
\addplot [black, very thick]
table {%
	3.775 2.485158075
	4.225 2.485158075
};
\addplot [black, very thick, mark=triangle*, mark size=3, mark options={solid}, only marks]
table {%
	4 2.473794409
};

\nextgroupplot[
tick align=outside,
tick pos=left,
width = 0.35\textwidth,
height = 0.3\textheight,
title={\textbf{ESTOI}},
x grid style={gray},
xmin=0.75, xmax=4.25,
xtick style={color=black},
xtick={1,2,3,4},
xticklabels={default, no DPCRNN, no skips, no styling},
xticklabel style = {rotate=12},
y grid style={gray},
ymajorgrids,
ymin=0.665781201690958, ymax=0.928858678713796,
ytick style={color=black}
]
\addplot [black, very thick]
table {%
	0.775 0.787708678005528
	1.225 0.787708678005528
	1.225 0.829479476386211
	0.775 0.829479476386211
	0.775 0.787708678005528
};
\addplot [black, very thick]
table {%
	1 0.787708678005528
	1 0.729173269595309
};
\addplot [black, very thick]
table {%
	1 0.829479476386211
	1 0.887588824877861
};
\addplot [black, very thick]
table {%
	0.8875 0.729173269595309
	1.1125 0.729173269595309
};
\addplot [black, very thick]
table {%
	0.8875 0.887588824877861
	1.1125 0.887588824877861
};
\addplot [black, very thick, mark=o, mark size=3, mark options={solid,fill opacity=0}, only marks]
table {%
	1 0.711530949843081
};
\addplot [black, very thick]
table {%
	1.775 0.791010350692494
	2.225 0.791010350692494
	2.225 0.822690092315092
	1.775 0.822690092315092
	1.775 0.791010350692494
};
\addplot [black, very thick]
table {%
	2 0.791010350692494
	2 0.748719769510897
};
\addplot [black, very thick]
table {%
	2 0.822690092315092
	2 0.864658751753708
};
\addplot [black, very thick]
table {%
	1.8875 0.748719769510897
	2.1125 0.748719769510897
};
\addplot [black, very thick]
table {%
	1.8875 0.864658751753708
	2.1125 0.864658751753708
};
\addplot [black, very thick, mark=o, mark size=3, mark options={solid,fill opacity=0}, only marks]
table {%
	2 0.73971846527511
	2 0.726313612561261
	2 0.720460700437427
	2 0.881207565854526
	2 0.878809252644077
};
\addplot [black, very thick]
table {%
	2.775 0.792258327111574
	3.225 0.792258327111574
	3.225 0.830001710914434
	2.775 0.830001710914434
	2.775 0.792258327111574
};
\addplot [black, very thick]
table {%
	3 0.792258327111574
	3 0.74179860633278
};
\addplot [black, very thick]
table {%
	3 0.830001710914434
	3 0.880717948529025
};
\addplot [black, very thick]
table {%
	2.8875 0.74179860633278
	3.1125 0.74179860633278
};
\addplot [black, very thick]
table {%
	2.8875 0.880717948529025
	3.1125 0.880717948529025
};
\addplot [black, very thick, mark=o, mark size=3, mark options={solid,fill opacity=0}, only marks]
table {%
	3 0.71608004477387
	3 0.731137831121098
	3 0.889182243674322
};
\addplot [black, very thick]
table {%
	3.775 0.768438848612764
	4.225 0.768438848612764
	4.225 0.802055559539234
	3.775 0.802055559539234
	3.775 0.768438848612764
};
\addplot [black, very thick]
table {%
	4 0.768438848612764
	4 0.72314910078917
};
\addplot [black, very thick]
table {%
	4 0.802055559539234
	4 0.835756827184315
};
\addplot [black, very thick]
table {%
	3.8875 0.72314910078917
	4.1125 0.72314910078917
};
\addplot [black, very thick]
table {%
	3.8875 0.835756827184315
	4.1125 0.835756827184315
};
\addplot [black, very thick, mark=o, mark size=3, mark options={solid,fill opacity=0}, only marks]
table {%
	4 0.675471647016168
	4 0.715155552848235
	4 0.857145237800036
	4 0.855475208543039
	4 0.870054352666864
};
\addplot [black, very thick]
table {%
	0.775 0.807631308663166
	1.225 0.807631308663166
};
\addplot [black, very thick, mark=triangle*, mark size=3, mark options={solid}, only marks]
table {%
	1 0.806632655868161
};
\addplot [black, very thick]
table {%
	1.775 0.80662588231481
	2.225 0.80662588231481
};
\addplot [black, very thick, mark=triangle*, mark size=3, mark options={solid}, only marks]
table {%
	2 0.805358998713194
};
\addplot [black, very thick]
table {%
	2.775 0.813349018822461
	3.225 0.813349018822461
};
\addplot [black, very thick, mark=triangle*, mark size=3, mark options={solid}, only marks]
table {%
	3 0.81089412535758
};
\addplot [black, very thick]
table {%
	3.775 0.788778372414008
	4.225 0.788778372414008
};
\addplot [black, very thick, mark=triangle*, mark size=3, mark options={solid}, only marks]
table {%
	4 0.785441825946041
};

\nextgroupplot[
tick align=outside,
tick pos=left,
width = 0.35\textwidth,
height = 0.3\textheight,
title={\textbf{NOMAD}},
x grid style={gray},
xmin=0.75, xmax=4.25,
xtick style={color=black},
xtick={1,2,3,4},
xticklabels={default, no DPCRNN, no skips, no styling},
xticklabel style = {rotate=12},
y grid style={gray},
ymajorgrids,
ymin=0.09405, ymax=1.47895,
ytick style={color=black}
]
\addplot [black, very thick]
table {%
	0.775 0.46425
	1.225 0.46425
	1.225 0.879
	0.775 0.879
	0.775 0.46425
};
\addplot [black, very thick]
table {%
	1 0.46425
	1 0.126
};
\addplot [black, very thick]
table {%
	1 0.879
	1 1.352
};
\addplot [black, very thick]
table {%
	0.8875 0.126
	1.1125 0.126
};
\addplot [black, very thick]
table {%
	0.8875 1.352
	1.1125 1.352
};
\addplot [black, very thick]
table {%
	1.775 0.571
	2.225 0.571
	2.225 0.94575
	1.775 0.94575
	1.775 0.571
};
\addplot [black, very thick]
table {%
	2 0.571
	2 0.297
};
\addplot [black, very thick]
table {%
	2 0.94575
	2 1.352
};
\addplot [black, very thick]
table {%
	1.8875 0.297
	2.1125 0.297
};
\addplot [black, very thick]
table {%
	1.8875 1.352
	2.1125 1.352
};
\addplot [black, very thick]
table {%
	2.775 0.64025
	3.225 0.64025
	3.225 0.92325
	2.775 0.92325
	2.775 0.64025
};
\addplot [black, very thick]
table {%
	3 0.64025
	3 0.278
};
\addplot [black, very thick]
table {%
	3 0.92325
	3 1.302
};
\addplot [black, very thick]
table {%
	2.8875 0.278
	3.1125 0.278
};
\addplot [black, very thick]
table {%
	2.8875 1.302
	3.1125 1.302
};
\addplot [black, very thick]
table {%
	3.775 0.64975
	4.225 0.64975
	4.225 1.08275
	3.775 1.08275
	3.775 0.64975
};
\addplot [black, very thick]
table {%
	4 0.64975
	4 0.317
};
\addplot [black, very thick]
table {%
	4 1.08275
	4 1.455
};
\addplot [black, very thick]
table {%
	3.8875 0.317
	4.1125 0.317
};
\addplot [black, very thick]
table {%
	3.8875 1.455
	4.1125 1.455
};
\addplot [black, very thick]
table {%
	0.775 0.765
	1.225 0.765
};
\addplot [black, very thick, mark=triangle*, mark size=3, mark options={solid}, only marks]
table {%
	1 0.697175
};
\addplot [black, very thick]
table {%
	1.775 0.7915
	2.225 0.7915
};
\addplot [black, very thick, mark=triangle*, mark size=3, mark options={solid}, only marks]
table {%
	2 0.76865
};
\addplot [black, very thick]
table {%
	2.775 0.7995
	3.225 0.7995
};
\addplot [black, very thick, mark=triangle*, mark size=3, mark options={solid}, only marks]
table {%
	3 0.7858
};
\addplot [black, very thick]
table {%
	3.775 0.961
	4.225 0.961
};
\addplot [black, very thick, mark=triangle*, mark size=3, mark options={solid}, only marks]
table {%
	4 0.882225
};
\end{groupplot}

\end{tikzpicture}\vspace{-20pt}
	\caption{Results of the ablation study at $1.5\,\mathrm{kbps}$ in terms of VisQol ($\uparrow$), ESTOI ($\uparrow$) and NOMAD ($\downarrow$). The black filled triangles mark the mean of the results.}
	\label{fig:ablation}
\end{figure*}
\begin{table}
	\begin{tabular}{lcccc}
		& Default & No DPCRNN & No skips & No styling \\
		\toprule
		Params. in millions & $3.61$ & $3.59$ & $3.33$ & $2.33$ \\
		$\mathrm{MMACs}$ & $343$ & $343$ & $330$ & $216$ \\
		\bottomrule
	\end{tabular}\vspace{3pt}
	\caption{Number of parameters and complexity in MMACs for all models of the ablation study.}
	\label{tab:ablation_cx}
\end{table}
\Revision{In the following, an ablation study is performed in order to show the impact of different building blocks of the proposed model. To this end, we train different variants of the model at $1.5\,\mathrm{kbps}$ with an $\mathtt{PSQ}$ realized with noise addition. In each of these trainings, we omit one of model building blocks and leave the other ones untouched.
}
More specifically, we remove the DPCRNN blocks (`no DPCRNN'), the TADEResBlocks for styling the \Revision{encoder} latent (`no skips') and the TADE blocks for styling the decoder (`no styling') individually for each model variant. The results in terms of VisQol, ESTOI and NOMAD can be found in Fig.~\ref{fig:ablation}. The impact of the DPCRNN and the styling of the latent is not significant in terms of \Revision{VISQOL} and ESTOI. However, in terms of NOMAD a clear improvement can be seen by adding these network components. The styling of the decoder has a more significant impact on the quality of the generated speech as can be seen \Revision{in the} evaluated objective measures. The influence of the evaluated building blocks may seem small at first sight, but \Revision{in relation} to the number of additional \ac{DNN} parameters and computational complexity shown in Tab.~\ref{tab:ablation_cx} it can be considered as a good \Revision{trade-off}: While the additional number of parameters and computational complexity stemming from the DPCRNNs is negligible (it even vanished by rounding in the values presented in Tab.~\ref{tab:ablation_cx}), the additional computational cost \Revision{from} the TADEResBlocks is \Revision{more visible but in the absolute minor. Styling the decoder with TADE layers adds significant complexity, which also results in a more pronounced improvement in speech quality.} 
\begin{figure}
\begin{tikzpicture}

\definecolor{darkgray176}{RGB}{176,176,176}
\definecolor{gray}{RGB}{128,128,128}
\definecolor{green}{RGB}{0,128,0}
\definecolor{orange}{RGB}{255,165,0}

\begin{axis}[
width=0.5\textwidth,
height=0.445\textheight,
tick align=outside,
tick pos=left,
axis x line*=middle,
axis y line=middle,
ymajorgrids,
x grid style={darkgray176},
xmin=0.0599999999999999, xmax=11.94,
xtick style={color=black},
xtick={1,2,3,4,5,7,8,9,11},
xticklabel style={rotate=45.0},
xticklabels={
  DIRECT,
  $Q=10$,
  $Q=17$,
  $Q=24$,
  $Q=31$,
  Dithered,
  Noise,
  ST,
  RVQ
},
y grid style={darkgray176},
ylabel={MOS},
ylabel style={at={(-0.15,0.5)}, rotate=90},
ymin=0, ymax=5.15,
ytick style={color=black}
]
\draw[draw=none,fill=gray] (axis cs:0.6,0) rectangle (axis cs:1.4,4.11818181818182);
\draw[draw=none,fill=lightgray] (axis cs:1.6,0) rectangle (axis cs:2.4,1.69090909090909);
\draw[draw=none,fill=lightgray] (axis cs:2.6,0) rectangle (axis cs:3.4,2.22727272727273);
\draw[draw=none,fill=lightgray] (axis cs:3.6,0) rectangle (axis cs:4.4,3.05454545454545);
\draw[draw=none,fill=lightgray] (axis cs:4.6,0) rectangle (axis cs:5.4,3.79090909090909);
\draw[draw=none,fill=green!50!white] (axis cs:6.6,0) rectangle (axis cs:7.4,3.14545454545455);
\draw[draw=none,fill=green] (axis cs:7.6,0) rectangle (axis cs:8.4,3.72727272727273);
\draw[draw=none,fill=green] (axis cs:8.6,0) rectangle (axis cs:9.4,3.61818181818182);
\draw[draw=none,fill=orange] (axis cs:10.6,0) rectangle (axis cs:11.4,3.81818181818182);
\path [draw=black, ultra thick]
(axis cs:1,3.97116014767865)
--(axis cs:1,4.26520348868499);

\path [draw=black, ultra thick]
(axis cs:2,1.52540587102058)
--(axis cs:2,1.8564123107976);

\path [draw=black, ultra thick]
(axis cs:3,2.02530048182961)
--(axis cs:3,2.42924497271584);

\path [draw=black, ultra thick]
(axis cs:4,2.84914831302275)
--(axis cs:4,3.25994259606816);

\path [draw=black, ultra thick]
(axis cs:5,3.64094512627355)
--(axis cs:5,3.94087305554463);

\path [draw=black, ultra thick]
(axis cs:7,2.96319984344502)
--(axis cs:7,3.32770924746407);

\path [draw=black, ultra thick]
(axis cs:8,3.56536357759298)
--(axis cs:8,3.88918187695247);

\path [draw=black, ultra thick]
(axis cs:9,3.45030216355811)
--(axis cs:9,3.78606147280553);

\path [draw=black, ultra thick]
(axis cs:11,3.67682648614186)
--(axis cs:11,3.95953715022177);

\node at (axis cs:3.5,0.75) [fill = white!50!lightgray, minimum width = 2cm] {MNRU};
\node at (axis cs:8,0.75) [fill = white!70!green, minimum width = 1.3cm] {SQ};

\node[rotate=35] at (axis cs:1,4.75) {$4.12$};
\node[rotate=35] at (axis cs:2,4.75) {$1.69$};
\node[rotate=35] at (axis cs:3,4.75) {$2.23$};
\node[rotate=35] at (axis cs:4,4.75) {$3.05$};
\node[rotate=35] at (axis cs:5,4.75) {$3.79$};
\node[rotate=35] at (axis cs:7,4.75) {$3.15$};
\node[rotate=35] at (axis cs:8,4.75) {$3.73$};
\node[rotate=35] at (axis cs:9,4.75) {$3.62$};
\node[rotate=35] at (axis cs:11,4.75) {$3.82$};

\end{axis}

\end{tikzpicture}
	\caption{P.808 ACR test results (including 22 listeners) comparing the different discussed quantization techniques \Revision{at $1.5\,\mathrm{kbps}$}. The values on top represent the mean and the black lines the \Revision{$95\%$} confidence intervals of the results.}
	\label{fig:LT_SQ_vs_VQ}
\end{figure}
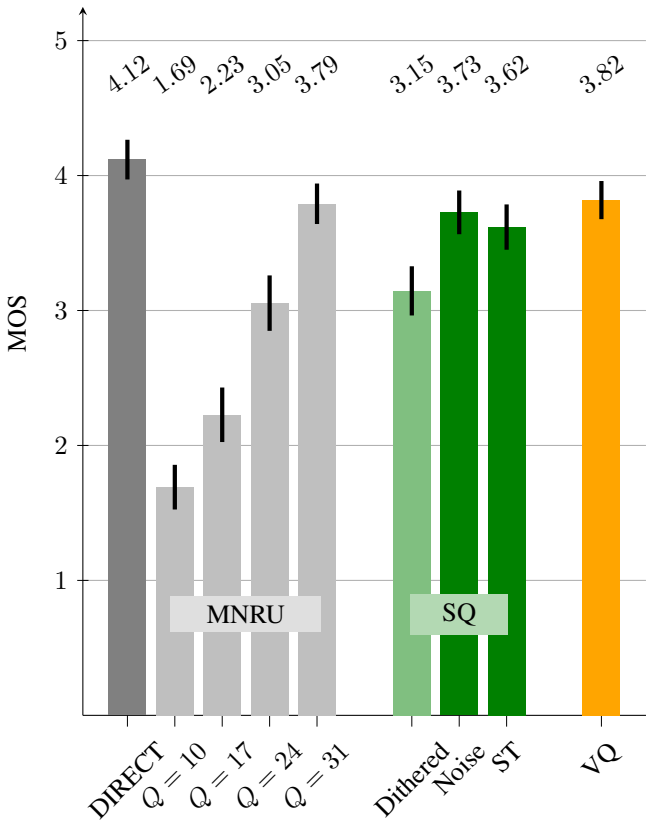
\begin{figure*}
\begin{tikzpicture}

\definecolor{cyan}{RGB}{0,255,255}
\definecolor{darkgray176}{RGB}{176,176,176}
\definecolor{gray}{RGB}{128,128,128}
\definecolor{green}{RGB}{0,128,0}
\definecolor{orange}{RGB}{255,165,0}

\begin{axis}[
width=\textwidth,
height=0.445\textheight,
tick align=outside,
tick pos=left,
axis x line*=middle,
axis y line=middle,
ymajorgrids,
x grid style={darkgray176},
xmin=0.06, xmax=24,
xtick style={color=black},
xtick={1,2,3,4,5,7,9,11,12,14,15,17,18,19,21,22,23},
xticklabel style={rotate=45.0},
xticklabels={
  DIRECT,
  $Q=10$,
  $Q=17$,
  $Q=24$,
  $Q=31$,
  $2.4\,\mathrm{kbps}$,
  $6\,\mathrm{kbps}$,
  $5.9\,\mathrm{kbps}$,
  $9.6\,\mathrm{kbps}$,
  $3.2\,\mathrm{kbps}$,
  $6\,\mathrm{kbps}$,
  $1\,\mathrm{kbps}$,
  $3\,\mathrm{kbps}$,
  $6\,\mathrm{kbps}$,
  $1\,\mathrm{kbps}$,
  $3\,\mathrm{kbps}$,
  $6\,\mathrm{kbps}$
},
y grid style={darkgray176},
ylabel={MOS},
ymajorgrids,
ymin=0, ymax=5.15,
ylabel style={at={(-0.075,0.5)}, rotate=90},
ytick={1,2,3,4,5},
yticklabels={1,2,3,4,5},
ytick style={color=black}
]

\draw[draw=none,fill=gray] (axis cs:0.6,0) rectangle (axis cs:1.4,4.12);
\draw[draw=none,fill=lightgray] (axis cs:1.6,0) rectangle (axis cs:2.4,1.9);
\draw[draw=none,fill=lightgray] (axis cs:2.6,0) rectangle (axis cs:3.4,2.67);
\draw[draw=none,fill=lightgray] (axis cs:3.6,0) rectangle (axis cs:4.4,3.31);
\draw[draw=none,fill=lightgray] (axis cs:4.6,0) rectangle (axis cs:5.4,3.79);

\draw[draw=none,fill=yellow!90!black] (axis cs:6.6,0) rectangle (axis cs:7.4,2.25);
\draw[draw=none,fill=blue] (axis cs:8.6,0) rectangle (axis cs:9.4,2.32);
\draw[draw=none,fill=cyan] (axis cs:10.6,0) rectangle (axis cs:11.4,3.79);
\draw[draw=none,fill=cyan] (axis cs:11.6,0) rectangle (axis cs:12.4,3.9125);
\draw[draw=none,fill=red] (axis cs:13.6,0) rectangle (axis cs:14.4,3.35);
\draw[draw=none,fill=red] (axis cs:14.6,0) rectangle (axis cs:15.4,3.77);
\draw[draw=none,fill=green] (axis cs:16.6,0) rectangle (axis cs:17.4,3.65);
\draw[draw=none,fill=green] (axis cs:17.6,0) rectangle (axis cs:18.4,3.6375);
\draw[draw=none,fill=green] (axis cs:18.6,0) rectangle (axis cs:19.4,3.875);
\draw[draw=none,fill=orange] (axis cs:20.6,0) rectangle (axis cs:21.4,2.6125);
\draw[draw=none,fill=orange] (axis cs:21.6,0) rectangle (axis cs:22.4,3.375);
\draw[draw=none,fill=orange] (axis cs:22.6,0) rectangle (axis cs:23.4,3.7125);

\path [draw=black, ultra thick]
(axis cs:1,3.97718752282674)
--(axis cs:1,4.26281247717326);

\path [draw=black, ultra thick]
(axis cs:2,1.74364608909738)
--(axis cs:2,2.05635391090262);

\path [draw=black, ultra thick]
(axis cs:3,2.49814612267535)
--(axis cs:3,2.84185387732465);

\path [draw=black, ultra thick]
(axis cs:4,3.15070793484859)
--(axis cs:4,3.46929206515141);

\path [draw=black, ultra thick]
(axis cs:5,3.62925296113132)
--(axis cs:5,3.95074703886868);

\path [draw=black, ultra thick]
(axis cs:7,2.04838841222977)
--(axis cs:7,2.45161158777023);

\path [draw=black, ultra thick]
(axis cs:9,2.12523134261645)
--(axis cs:9,2.51476865738355);

\path [draw=black, ultra thick]
(axis cs:11,3.63666582872429)
--(axis cs:11,3.94333417127571);

\path [draw=black, ultra thick]
(axis cs:12,3.70861470865873)
--(axis cs:12,4.11638529134127);

\path [draw=black, ultra thick]
(axis cs:14,3.15245545286881)
--(axis cs:14,3.54754454713119);

\path [draw=black, ultra thick]
(axis cs:15,3.59412873854418)
--(axis cs:15,3.94587126145582);

\path [draw=black, ultra thick]
(axis cs:17,3.44639439961944)
--(axis cs:17,3.85360560038056);

\path [draw=black, ultra thick]
(axis cs:18,3.43346569691485)
--(axis cs:18,3.84153430308515);

\path [draw=black, ultra thick]
(axis cs:19,3.67357045708364)
--(axis cs:19,4.07642954291636);

\path [draw=black, ultra thick]
(axis cs:21,2.38521275771322)
--(axis cs:21,2.83978724228678);

\path [draw=black, ultra thick]
(axis cs:22,3.16762292034905)
--(axis cs:22,3.58237707965095);

\path [draw=black, ultra thick]
(axis cs:23,3.51161815457144)
--(axis cs:23,3.91338184542856);

\node at (axis cs:3.5,0.75) [fill = white!50!lightgray, minimum width = 2cm] {MNRU};
\node at (axis cs:7,0.75) [fill = white!70!yellow, minimum width = 1.3cm] {Codec2};
\node at (axis cs:9,0.75) [fill = white!70!blue, minimum width = 1.3cm] {Opus};
\node at (axis cs:11.5,0.75) [fill = white!70!cyan, minimum width = 1.3cm] {EVS};
\node at (axis cs:14.5,0.75) [fill = white!70!red, minimum width = 1.3cm] {Lyra V2};
\node at (axis cs:18,0.75) [fill = white!70!green, minimum width = 1.3cm] {Proposed (SQ)};
\node at (axis cs:22,0.75) [fill = white!70!orange, minimum width = 1.3cm] {FreqCodec};

\node[rotate=35] at (axis cs:1,4.75) {$4.12$};
\node[rotate=35] at (axis cs:2,4.75) {$1.90$};
\node[rotate=35] at (axis cs:3,4.75) {$2.67$};
\node[rotate=35] at (axis cs:4,4.75) {$3.31$};
\node[rotate=35] at (axis cs:5,4.75) {$3.79$};

\node[rotate=35] at (axis cs:7,4.75) {$2.25$};
\node[rotate=35] at (axis cs:9,4.75) {$2.32$};

\node[rotate=35] at (axis cs:11,4.75) {$3.79$};
\node[rotate=35] at (axis cs:12,4.75) {$3.91$};

\node[rotate=35] at (axis cs:14,4.75) {$3.35$};
\node[rotate=35] at (axis cs:15,4.75) {$3.77$};

\node[rotate=35] at (axis cs:17,4.75) {$3.65$};
\node[rotate=35] at (axis cs:18,4.75) {$3.64$};
\node[rotate=35] at (axis cs:19,4.75) {$3.88$};

\node[rotate=35] at (axis cs:21,4.75) {$2.61$};
\node[rotate=35] at (axis cs:22,4.75) {$3.38$};
\node[rotate=35] at (axis cs:23,4.75) {$3.71$};

\end{axis}

\end{tikzpicture}\vspace{-10pt}
	\caption{\Revision{P.808 ACR test results (including 20 listeners) comparing the proposed \ac{STFT}-based neural codec using \ac{SQ} with FreqCodec using \ac{RVQ}, Codec2, Opus, EVS and Lyra V2 a neural codec based on soundstream \cite{zeghidour_soundstream_2021} at various bitrates. The values on top represent the mean and the black lines the $95\%$ confidence intervals of the results.}}
	\label{fig:LT_baselines}
\end{figure*}
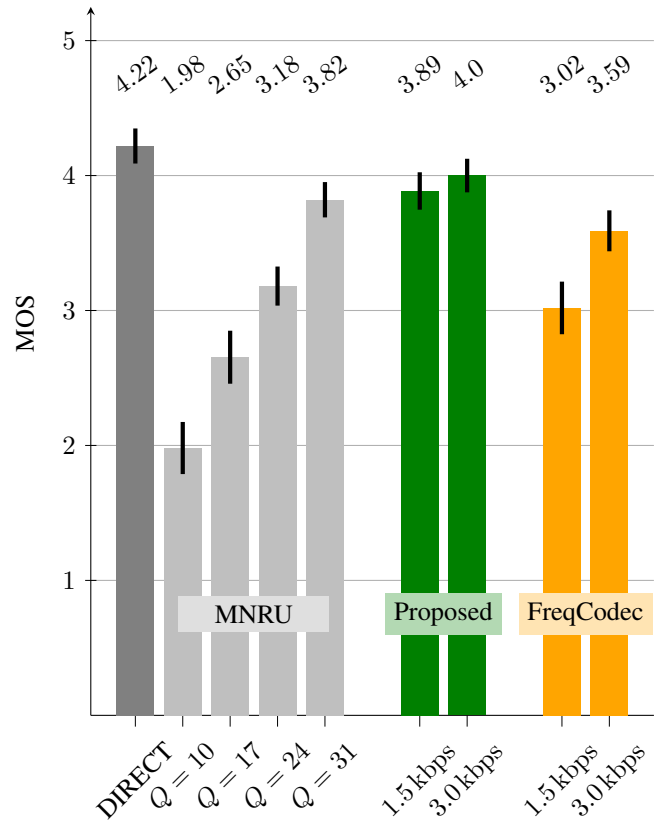
\begin{figure}\vspace{-12pt}
\begin{tikzpicture}

\definecolor{darkgray176}{RGB}{176,176,176}
\definecolor{gray}{RGB}{128,128,128}
\definecolor{green}{RGB}{0,128,0}
\definecolor{orange}{RGB}{255,165,0}

\begin{axis}[
width=0.5\textwidth,
height=0.445\textheight,
tick align=outside,
tick pos=left,
axis x line*=middle,
axis y line=middle,
ymajorgrids,
x grid style={darkgray176},
xmin=0.0599999999999999, xmax=11.94,
xtick style={color=black},
xtick={1,2,3,4,5,7,8,10,11},
xticklabel style={rotate=45.0},
xticklabels={
  DIRECT,
  $Q=10$,
  $Q=17$,
  $Q=24$,
  $Q=31$,
  SQ,
  RVQ,
  SQ,
  RVQ
},
y grid style={darkgray176},
ylabel={MOS},
ymajorgrids,
ymin=0, ymax=5.15,
ytick={1,2,3,4,5},
yticklabels={1,2,3,4,5},
ylabel style={at={(-0.15,0.5)}, rotate=90},
ymin=0, ymax=5.25,
ytick style={color=black}
]

\draw[draw=none,fill=gray] (axis cs:0.6,0) rectangle (axis cs:1.4,4.13684210526316);
\draw[draw=none,fill=lightgray] (axis cs:1.6,0) rectangle (axis cs:2.4,2.33684210526316);
\draw[draw=none,fill=lightgray] (axis cs:2.6,0) rectangle (axis cs:3.4,2.84210526315789);
\draw[draw=none,fill=lightgray] (axis cs:3.6,0) rectangle (axis cs:4.4,3.44210526315789);
\draw[draw=none,fill=lightgray] (axis cs:4.6,0) rectangle (axis cs:5.4,3.81052631578947);
\draw[draw=none,fill=green] (axis cs:6.6,0) rectangle (axis cs:7.4,3.76842105263158);
\draw[draw=none,fill=green] (axis cs:7.6,0) rectangle (axis cs:8.4,3.69473684210526);
\draw[draw=none,fill=orange] (axis cs:9.6,0) rectangle (axis cs:10.4,3.11578947368421);
\draw[draw=none,fill=orange] (axis cs:10.6,0) rectangle (axis cs:11.4,3.33684210526316);
\path [draw=black, ultra thick]
(axis cs:1,3.97455622400647)
--(axis cs:1,4.29912798651985);

\path [draw=black, ultra thick]
(axis cs:2,2.13233401114771)
--(axis cs:2,2.5413501993786);

\path [draw=black, ultra thick]
(axis cs:3,2.64684018256139)
--(axis cs:3,3.0373703437544);

\path [draw=black, ultra thick]
(axis cs:4,3.274331501422)
--(axis cs:4,3.60987902489379);

\path [draw=black, ultra thick]
(axis cs:5,3.66300308243591)
--(axis cs:5,3.95804954914303);

\path [draw=black, ultra thick]
(axis cs:7,3.61755908025765)
--(axis cs:7,3.91928302500551);

\path [draw=black, ultra thick]
(axis cs:8,3.51856333932333)
--(axis cs:8,3.87091034488719);

\path [draw=black, ultra thick]
(axis cs:10,2.93059364961788)
--(axis cs:10,3.30098529775054);

\path [draw=black, ultra thick]
(axis cs:11,3.1476308598649)
--(axis cs:11,3.52605335066142);

\node at (axis cs:3.5,0.75) [fill = white!50!lightgray, minimum width = 2cm] {MNRU};
\node at (axis cs:7.5,0.75) [fill = white!70!green, minimum width = 1.3cm] {Proposed};
\node at (axis cs:10.5,0.75) [fill = white!70!orange, minimum width = 1.3cm] {FreqCodec};

\node[rotate=35] at (axis cs:1,4.75) {$4.14$};
\node[rotate=35] at (axis cs:2,4.75) {$2.34$};
\node[rotate=35] at (axis cs:3,4.75) {$2.84$};
\node[rotate=35] at (axis cs:4,4.75) {$3.44$};
\node[rotate=35] at (axis cs:5,4.75) {$3.81$};
\node[rotate=35] at (axis cs:7,4.75) {$3.77$};
\node[rotate=35] at (axis cs:8,4.75) {$3.69$};
\node[rotate=35] at (axis cs:10,4.75) {$3.12$};
\node[rotate=35] at (axis cs:11,4.75) {$3.34$};

\end{axis}

\end{tikzpicture}\vspace{-10pt}
	\caption{\Revision{P.808 ACR test results (including 19 listeners) comparing the proposed \ac{STFT}-based codec with FreqCodec \cite{du_funcodec_2023} each evaluated for both \ac{SQ} and \ac{RVQ} at $1.5\,\mathrm{kbps}$. The values on top represent the mean and the black lines the $95\%$ confidence intervals of the results.}}
	\label{fig:LT_sq_vq}
\end{figure}
\subsection{Comparison of Quantizers}
In the following, we compare the different quantization approaches discussed in this paper, i.e., Noise \ac{SQ}, ST \ac{SQ} and \ac{VQ}. All models are trained end-to-end with the mentioned \Revision{quantizers}. For completeness, we also evaluate dithered quantization (cf. \eqref{eq:dithered_SQ}) on the model trained with Noise \ac{SQ}. Note that the latent of the Noise \ac{SQ} model and the latent of the same model evaluated with dithered quantization may follow the same distribution (see Sec.~\ref{sec:SQ_dithered} for this discussion) but the particular realizations during evaluation may differ causing a training-test mismatch \Revision{for the dithered version}.

The results of the corresponding listening test comprising $22$ listeners (corresponding to 88 votes per condition) can be seen in Fig.~\ref{fig:LT_SQ_vs_VQ}: The performance of Noise \ac{SQ}, ST \ac{SQ} and \ac{VQ} does not show significant differences, i.e., despite the simpler implementation, the \ac{SQ}-based methods perform as well as the \ac{VQ} approach. Dithered quantization performs worse than the other approaches which is to be attributed to the mentioned \Revision{training-test} mismatch. However, the performance does not drop too much suggesting that the latent representation learned with Noise \ac{SQ} is \Revision{smooth enough to allow random perturbations without impacting significantly the reconstruction quality}.
%
\subsection{\Revision{Comparison of Quantizers for Different Neural Codecs}}
\label{sec:SQ_VQ_STFTvsFreqCodec}
\Revision{In this section, we investigate the behavior of the discussed quantization techniques for another state-of-the-art model. As a method from the literature that also operates in the \ac{STFT} domain, we choose FreqCodec \cite{du_funcodec_2023} which was developed based on \ac{RVQ}. For a reasonable comparison with our method, we use the model termed M6 in the paper which has $0.52$ million parameters whose complexity is $340\,\mathrm{MMACs}$. Hence, the complexity of both models matches almost perfectly (cf. Tab.~\ref{tab:ablation_cx}), although our model has more parameters. However, FreqCodec is not causal, i.e., introduces significant algorithmic latency compared to our causal model. For a fair comparison, we trained the FreqCodec model with the same data set and the code and recipes provided by the authors. Informal listening showed that this provides the same performance or even a slight improvement relative to the available pretrained models. For this evaluation, we trained the proposed model and FreqCodec both using \ac{VQ} and \ac{SQ} at $1.5\,\mathrm{kbps}$. As the listening test in Fig.~\ref{fig:LT_SQ_vs_VQ} showed no significant difference between \ac{SQ} realizations, we pick the \ac{SQ} realization by noise addition as training scheme for the \ac{SQ}-based models.} 

\Revision{The results of the corresponding listening test comprising 19 listeners (76 votes per condition) is shown in Fig.~\ref{fig:LT_sq_vq}. For the proposed neural codec as well as for FreqCodec, both \ac{SQ} and \ac{VQ} perform comparably. However, the proposed audio codec significantly outperforms FreqCodec for both quantizers.}
%
\subsection{\Revision{Comparison with Baselines}}
%
%
\Revision{For benchmarking the proposed neural codec, we compare its performance against baseline models following various approaches across a variety of data rates. The set of baseline methods include classical signal processing-based codecs, i.e., Codec2 \cite{rowe_codec_2024} at $2.4\,\mathrm{kbps}$, a parametric codec, the open source codec Opus \cite{rfc6716} at $6\,\mathrm{kbps}$, and EVS \cite{evs}, a widely adopted codec for mobile communications, at $5.9\,\mathrm{kbps}$ and $9.6\,\mathrm{kbps}$. For comparison with end-to-end trained neural codecs, we include Lyra V2 \cite{noauthor_lyra_2022}, a neural codec based on soundstream \cite{zeghidour_soundstream_2021}, at $3.2\,\mathrm{kbps}$ and $6\,\mathrm{kbps}$ in the set of baselines. Furthermore, we evaluate FreqCodec using \ac{VQ} and the proposed neural codec using \ac{SQ} (training with noise addition) both at $1\,\mathrm{kbps}$, $3\,\mathrm{kbps}$ and $6\,\mathrm{kbps}$. The results of the corresponding listening test are depicted in Fig.~\ref{fig:LT_baselines}.}

\Revision{The signal processing-based codecs Codec2 and Opus show significantly lower performance than all other competing methods. EVS provides the overall best quality, however, also at the highest bitrates among the tested methods. Lyra V2 achieves similar performance as FreqCodec. Finally, the proposed method outperforms all baseline methods at comparable bitrates, where its superiority is especially pronounced at lower bitrates, see, e.g., its performance at $1\,\mathrm{kbps}$ compared to FreqCodec.}
\section{Conclusion}
\label{sec:conclusion}
In this paper, we proposed the use of simple \ac{SQ}-based quantization \Revision{in} neural audio coding and \Revision{carried out} in-depth analysis of \Revision{different realizations of the technique}. In particular, we showed relations to \ac{VQ}, \acp{VAE}, regularization and dithered quantization. Furthermore, \Revision{and to meet the constraints of real-time communication, we proposed a novel neural speech coding model taking advantage of \ac{SQ} and being able to achieve high speech quality at low bitrates and low computational complexity.} \Revision{Through this, we confirm} the potential of \ac{STFT}-based neural audio codecs, which are underrepresented in the literature so far. In experiments we showed that the proposed simple \Revision{\ac{SQ}-based techniques} perform \Revision{on par with} \ac{VQ} and that the proposed \Revision{neural speech codec} outperforms a recent baseline of same complexity by a large margin.




\bibliographystyle{IEEEtran}
\bibliography{literature}
%


%
%
%
%


\end{document}